\documentclass[final,3p,times]{elsarticle}
\usepackage{amsmath, amsfonts, amssymb}
\usepackage{graphicx}
\usepackage{natbib}
\usepackage{bbm}
\usepackage{color}

\newcommand{\ket}[1]{| #1 \rangle}
\newcommand{\bra}[1]{\langle #1 |}

\newcommand{\ex}[1]{\langle #1 \rangle}

\newcommand{\beq}{\begin{eqnarray}}
\newcommand{\eeq}{\end{eqnarray}}

\journal{Physics Reports}

\begin{document}

\begin{frontmatter}
\title{Functional quantum biology in photosynthesis and magnetoreception}
\author[riken]{Neill Lambert\footnote{nwlambert@riken.jp}}
\author[CK]{Yueh-Nan Chen\footnote{yuehnan@mail.ncku.edu.tw}}
\author[TP]{Yuan-Chung Cheng\footnote{yuanchung@ntu.edu.tw}}
\author[CKES]{Che-Ming Li}
\author[CK]{Guang-Yin Chen}
\author[riken,Mich]{Franco Nori}
\address[riken]{Advanced Science Institute, RIKEN, Saitama 351-0198, Japan}
\address[CK]{Department of Physics and National Center for Theoretical
Sciences, National Cheng-Kung University, Tainan 701, Taiwan}
\address[TP]{Department of Chemistry and Center for Quantum Science and Engineering, National Taiwan University, Taipei City 106, Taiwan}
\address[CKES]{Department of Engineering Science and Supercomputing Research Center,
National Cheng Kung University, Tainan 701, Taiwan}
\address[Mich]{Physics Department, The University of Michigan, Ann Arbor, Michigan, 48109-1040, USA}

\begin{abstract}
Is there a functional role for quantum mechanics or coherent
quantum effects in biological processes?  While this question is
as old as quantum theory, only recently have measurements on
biological systems on ultra-fast time-scales shed light on a
possible answer. In this review we give an overview of the two
main candidates for biological systems which may harness such
functional quantum effects: photosynthesis and magnetoreception.
We discuss some of the latest evidence both for and against room
temperature quantum coherence, and consider whether there is truly
a functional role for coherence in these biological mechanisms.
Finally, we give a brief overview of some
 more speculative examples of functional quantum biology including
 the sense of smell, long-range quantum tunneling in proteins,
biological photoreceptors, and the flow of ions across a cell
membrane.
\end{abstract}


\end{frontmatter}

\section{Introduction}

The role of quantum mechanics in biological processes has a long history %
\cite{schrodinger,davies,LonguetHiggins:1962p85306}. After all, biological molecules are
constructed of atoms, and bound together by forces of a `quantum
origin'. However the role of quantum coherence (and more recently
entanglement) on time or energy scales which play
a direct role in biological function has remained controversial. %
Recent
experiments~\cite{EngelNature07,LeeScience07,ambientTemperature,PNAS,Panitchayangkoon:2011cs}
on photosynthetic `Light Harvesting Complexes' (LHC) and their
constituents (e.g., the Fenna--Matthews--Olson (FMO)
pigment-protein complex in green sulfur bacteria) have suggested
that quantum coherence may play a role in one of the most
fundamental and important of biological processes: energy
transport and energy conversion.

LHC complexes are arrangements of pigments (most importantly
chlorophylls) and protein molecules which function as a light
gathering `antenna' to absorb photons and become electronically
excited \cite{vanAmerongen:2000tb,Blankenship}. This excitation is
then passed to a reaction center, where it is converted into
useful chemical energy. One of the most interesting features of
this biological machine is the highly efficient transfer mechanism
which takes the electronic excitation through the LHC to the
reaction center with almost unity quantum yield. In green sulfur
bacteria, part of this chain connecting the antenna (which in
green bacteria is a large chlorosome) to the reaction center is
composed of the Fenna--Matthews--Olson (FMO) pigment-protein
complexes \cite{Olson:2004p29865}; see Fig. 1 for an explanatory schematic.  It is this
complex which has been most widely studied, and will form part of the focus
of this review.

In photosynthetic light harvesting quantum coherence can have
different meanings in different circumstances. Quantum coherence
can refer to quantum superpositions of localized molecular
excitations that occur naturally because electronic couplings
between molecular excitations lead to delocalized eigenstates
(a.k.a. excitons). Sometimes, coherence in light harvesting
alludes to the coherent wave-like dynamics of energy transfer,
which actually reflects the superposition of excitonic
eigenstates. In the former case, the coherence is represented in
the molecular site basis, whereas in the latter case, the
coherence is represented in the delocalized exciton basis. Note
that the site basis and the exciton basis are special because they
are related to the spatial arrangement of chromophores and energy
eigenstates of the Hamiltonian, respectively. Although both types
of coherence effects play important roles in photosynthetic light
harvesting, they must be discussed separately. Quantum coherence
manifested in the delocalized eigenstates of photoexcitations in
photosynthetic complexes plays a fundamental role in spectral
properties, energy tuning, and energy transfer dynamics of
photosynthetic light harvesting
\cite{Yen:2011p99898,Fleming:2011p99897,Scholes:2011kx}. The
effects of excitonic coherence are more difficult to analyze and
have been the subject of intensive research in the past few years.  Spectroscopic measurements (termed two-dimensional electronic spectroscopy) on the FMO complex at liquid-nitrogen (77K)~\cite%
{EngelNature07} and room temperatures~\cite{,PNAS} have shown
time-dependent oscillations, presumed to be quantum beating, in
the amplitudes of the spectral signals, which matches the
predictions of quantum theory. Inspired by these observations, it
has been proposed that a intricate interplay of quantum coherent
excitation transfer and environmental dephasing help increase the
efficiency of the energy transport process ~\cite%
{Plenio08,Lloyd08,Caruso09,cheng2009dynamics,LeeCheng09}.

Apart from the example of photosynthesis it is also believed that
 other functional biological mechanisms may display, or rely upon,
quantum effects. For example, the second part of this review will focus on the singlet and triplet states of
spatially-separated electron spins in radical pairs
(donor-acceptor molecules), whichh are
hypothesized to play a role in avian magnetoreception \cite%
{johnsen2005physics,ritz2010photoreceptor,Johnsen2008}. If true,
this would represent a functional piece of biological ``quantum
hardware'' that could not function in a classical
world~\cite{vedral11}. More speculative examples we will describe
more briefly include long-range quantum tunneling in
proteins~\cite{Gray:2003p83392,Stuchebrukhov:2003p83910}, the
sense of smell~\cite{turin96}, biological
photoreceptors~\cite{Schoenlein:1991to,vanderHorst:2004cc},
and the flow of ions across a cell membrane~\cite{Vaziri10}.

\subsection{Functionality defined}

Our goal here is to give a brief overview of these different
biological mechanisms, summarize some of the theoretical models,
and highlight some of the experimental and theoretical evidence
both for and against a functional role for quantum effects in
biological systems. By ``functional'' we imply a role where the
presence of coherent quantum dynamics achieves something either
more efficiently, or otherwise impossible, than
could be achieved by a classical mechanism alone. 
This concept of nature taking advantage of quantum mechanics is an
inspiring notion, but the evidence for and against it must be
examined carefully. We thus discuss only experimentally-verifiable systems in this review. This topic is growing at a
phenomenal pace, and we can only summarize the main points for
each system we consider.

\subsubsection{Photosynthesis}

In the example of energy transfer in photosynthetic complexes,
most of the quantum effects observed so far have been at
liquid-Nitrogen temperatures ($77$ K) \cite{EngelNature07}, though
recent evidence has arisen of persistence quantum coherence at
room temperature, both in FMO \cite{PNAS} and in other components
of certain light-harvesting complexes \cite{ambientTemperature}. To compliment this a
variety of theoretical techniques have been used to show that a
quantum model of the FMO complex implies a high transport efficiency even at room temperature %
\cite{Plenio08,Lloyd08,Caruso09,Plenio10,Plenio102}. As mentioned,
this amalgamation of experimental observation and theoretical
modeling suggests that a combination of quantum effects and
environmental `stimulation' leads to a high efficiency in the rate
of energy transport in the FMO complex. However, alternative
points of view still exist in the literature~\cite{Mancal:2010dj}, and more work
remains to be done to analyze to what extent this environment-assisted transport is really `quantum'. For example, in principle
one must eliminate all possible classical models~\cite{Briggs11}
and purely thermal/environmental effects~\cite{Olbrich11} that
could produce similar efficiencies before unambiguously stating
that high efficiency transport itself is an example of functional
quantum biology. In addition, it is not yet clear if the dynamics
observed in-vitro (in experiment) also occur in-vivo (in nature),
due to uncertainty about the nature of the energy transport from
the antenna (chlorosome and baseplate) to the FMO complex.  We
will attempt to summarize all of these issues in this article.

\subsection{Avian magnetoreception}

In contrast to photosynthesis, the functionality of the proposed
radical pair model for avian magnetoreception is entirely
dependent on quantum mechanics~\cite{vedral11}. In this sense it
is different from the possibly more ambiguous role of quantum
mechanics in photosynthesis. For example, the proposed mechanism
of radical-pair magnetoreception cannot function at all without a
large degree of spin-spin entanglement and
coherence~\cite{vedral11}: if environmental noise is too strong
(an effective classical limit of the model) then birds could not
navigate via this proposed mechanism. In addition, while ubiquitous in molecular and atomic physics, there is
essentially also no classical analogue to the singlet/triplet
states of two coupled spins, which is the fundamental element of
this model which acts as a sensor of the Earth's magnetic field.

The open question in this case then is whether such a proposed radical pair
really exists and functions as predicted. There is some suggestion
it may reside in cryptochromes (a light-sensitive cell found in
the eye).  However, experiments on a range of possible
radical-pair molecules in laboratory conditions have not yet found
a radical-pair molecule with the desired properties. Furthermore,
evidence must also be found for whether birds really have the
biological circuitry in place to process the signal that
radical-pair magnetoreception would generate.  Later we will
summarize the details of this mechanism, explain why it is a
strong contender for functional quantum biology, and outline the
evidence supporting it as a mechanism for
magnetoreception.%

\section{Photosynthesis}

\begin{table*}[th]
{\small \hfill {}
\begin{tabular}{ccc}
\hline
\textbf{Biological system} &  & \textbf{Results} \\ \hline\hline
\textbf{Photosynthesis } &  &  \\
& Cyrogenic temperature quantum coherence & \cite{LeeScience07,EngelNature07}\\
& Ambient/room temperature quantum coherence (FMO) & \cite{PNAS}
\\ & Ambient/room temperature quantum
coherence (Algae) & \cite{ambientTemperature} \\
& Environment assisted transport & \cite{Plenio08, Lloyd08,Akihito09,LeeCheng09} \\
& Entanglement, Leggett-Garg & \cite{Sarovar10,Wilde10} \\
& Alternative views & \cite{Tiersch11,Briggs11}  \\
\hline
\textbf{Radical Pair Magnetoreception} &  &  \\
& Early proposals and evidence & \cite{Schulten, wilt72} \\
& Mathematical models &  \cite{Schulten}, \cite{Ritz00} \\
& Indirect evidence (light dependance, magnetic field) & \cite{Wiltschko00,WiltLight,Wiltschko02a,Ritz04, Wiltschko07,Ritz09} \\
& Experiments on Radical pairs &  \cite{Steiner89,woodward01,Christopher07,Maeda08} \\
& Monarch butterflies, Flys, Humans & \cite{Foley10, Gegear}  \\
\hline
\textbf{Other examples} &  &  \\
& Olfaction & \cite{turin96,Brookes07}  \\
&Ion channels & \cite{Vaziri10}\\
&Vision & \cite{Schoenlein:1991to,Polli:2010bk} \\
&Long-range electron transfer & \cite{Gray:2003p83392,Stuchebrukhov:2003p83910} \\
&Enzyme catalysis & \cite{Nagel:2006p82823,Allemann:2009uq} \\
\hline
\end{tabular}
} \hfill {} \caption{Summary of a selection of the main
experimental and theoretical works on functional quantum biology.}
\end{table*}

Photosynthesis is one of the most important photochemical
processes on Earth. As outlined in the introduction, the primary
photosynthetic apparatus (which is typically called a
photosynthetic unit, or PSU) can be roughly divided into several
parts: a light-harvesting complex antenna, and a reaction center
(RC), as shown in Fig. 1 for the example of green sulfur bacteria.
The function of the FMO complex in green sulfur bacteria is to transport
excitations from the sunlight-harvesting LHC antenna (chlorosome)
to the reaction center (RC), where it initiates a
charge-separation process that generates chemical potential for
biochemistry.

The FMO complex consists of eight
bacteriochlorophyll-a (BChl-a) molecules which are bound to a
surrounding protein scaffolding. Note that the eighth BChl was
not discovered in structural models of the FMO complexes until
recently, therefore most studies on the FMO complex so far only
considered a model with seven BChls. As we will describe later, in a
simplified picture, each BChl molecule can be excited from its
ground state into its first singlet excited state, forming a
molecular exciton. Electronic couplings between molecular excitons
then enable excitation energy transfer between BChl molecules.

After a photon is caught by the chlorosome antenna, it is
transferred through the baseplate to the FMO complex (see Fig. 1).
This excitation is then passed from one BChl molecule to the
next, until it reaches the molecule, or site, closest to the
reaction center.  It then irreversibly enters the reaction center,
and ignites the charge-separation process. In general terms, this
apparatus functions in a similar way across a broad range of
photosynthetic organisms, though the size and configuration of the
antenna and component systems can differ
greatly~\cite{Cogdell:2008p59896} (e.g., see Hu et al \cite{Hu02}
for a review of the PSU in purple bacteria).

The most well-known way to describe this excitation-transfer
process is the F\"{o}rster model \cite{Forster48}. In this model
the dipole-transition coupling between two chromophores (sites) gives
rise to rates for excitation (exciton) hopping between interacting
sites. The F\"{o}rster model is a successful one, but it
neglects quantum coherences between different sites. Experiments
have shown that the exciton can move coherently among several
chromophores, leading to electronic quantum beating~\cite{Savikhin97},
and perhaps also coherent collective phenomena like the well-known
optical example of superradiance~\cite{Monshouwer:1997p32861,Zhao:1999p3586,Leupold96,Grondelle06}.

In addition, some other components of LHCs in other species
suggest that coherence and long-range entanglement may exist in
larger structures.  For example, the LHC of certain types of
purple bacteria contain two types of ring-like antenna, LH1 and
LH2 \cite{Hu02}, which are substantially larger than the FMO
complex. Spectroscopic studies on these systems have suggested that
photoexcitations may delocalize among 4-5 BChl sites in the strongly coupled rings~\cite{Monshouwer:1997p32861,Zhao:1999p3586,Grondelle06}.  

Another interesting issue is that the energy transport through the
FMO complex is very fast (100's of femto-seconds), and the
efficiency of many LHC systems is very high. The quantum yield of
light-to-charge conversion in some photosynthetic units
\cite{Blankenship} can be up to 95\%, and explaining how the
excitations navigate the energy landscape of the LHC so
successfully is non-trivial.  In addition, the need to navigate
this landscape quickly must be understood in terms of losses that
occur during the transport through the LHC system and its
components, and not in terms of the overall energy processing rate
of the entire PSU (LHC and RC systems), which is relatively slow
in comparison. For example, the `reset time' or `turnover' rate of
the RC in purple bacteria~\cite{Hu02} is of the order of $1$ kHz,
and the energy absorption rate of a single BChl in an antenna in
bright conditions is $10$ Hz.  It is thought that the large number
of BChl in the antenna in totality provide enough energy to
optimally use this $1$ kHz turnover rate of the RC.  Thus there is
a large separation of time-scales.  The fast transport through the
FMO complex, for example, is needed to ``beat'' the rate at which
excitations are lost due to fluorescence relaxation, not because
of the overall ``clock rate'' of the light harvesting complex.
Also, one should note that
 the ``energy'' efficiency of photosynthetic bio-mass as, e.g., a fuel
 source, is relatively low even compared with photovoltaic
efficiency~\cite{Barber09, Blankenship11}, because the down-stream
biochemical reactions that turn chemical potentials into biomass have
extremely low energy efficiency.

To gain a deeper understanding of these quantum phenomena,
powerful spectroscopic tools have been developed, which provide opportunities to
study the dynamics of the excitations in these pigment-protein
complexes at the time scales of femto-seconds
\cite{Grondelle06,Cheng:2008p51139}. For example, two-dimensional electronic
spectroscopy is a powerful
technique that probes the couplings and the dynamics of
energy flow on a two-dimensional (2D) map in the frequency domain that
allows direct observation of ``coherence''
between electronic excitations.  
A full understanding of exactly how much information can be extracted
with such techniques
is still being discussed and researched \cite{Cheng07,Guzik11,Guzik112}.

\begin{figure}[h]
\includegraphics[width=0.5\columnwidth]{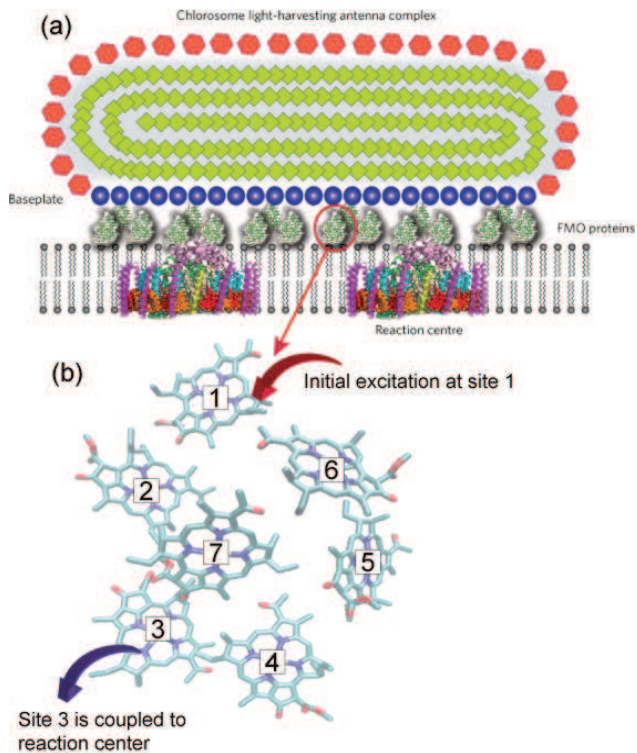}
\caption{{} (Color online) (a) Schematic diagram (from
\cite{ScholesNV}) of roughly 200,000 BChl-c molecules (green
rhombuses) encased in a protein–lipid structure form the
chlorosome antenna. This is much larger than antennas seen in
other species, possibly due to the low-light conditions in which
these bacteria thrive. This antenna harvests sunlight, and then
the excitation energy is transferred (via baseplate proteins,
represented by the blue circles) to several FMO complexes. These
act as a wire to transport the energy to reaction centers
 in the membrane. The reaction center shown here
is from Chloroflexus aurantiacus \cite{xin}.  The red hexagons
represent the chlorosome envelope.  Below this, in (b), there is a
rendering of the structure of one of the FMO pigment protein
complexes. The function of FMO in green sulfur bacteria is to
transport excitations from the sunlight-harvesting LHC antenna
(chlorosome) to the RC where it serves to initiate a
charge-separation process. The FMO complex consists of eight
bacteriochlorophyll a (BChl a) which are bound to a surrounding
protein scaffolding. The excitation received from the antenna
(here at molecule, or site, $1$) is then passed from one BChl
molecule to the next, until it reaches the molecule, or site,
closest to the reaction center (here assumed to be site $3$). It
then irreversibly enters the reaction center, and ignites the
charge separation process. The site labels here correspond to the
states $\ket{j}$ is the Hamiltonian in Eq.~(\ref{Hamiltonian}) and
Eq. (\ref{Hamiltonian2}). \label{FMO}}
\end{figure}

\subsection{Experimental signatures of quantumness}

\begin{figure}[h]
\includegraphics[width=0.8\columnwidth]{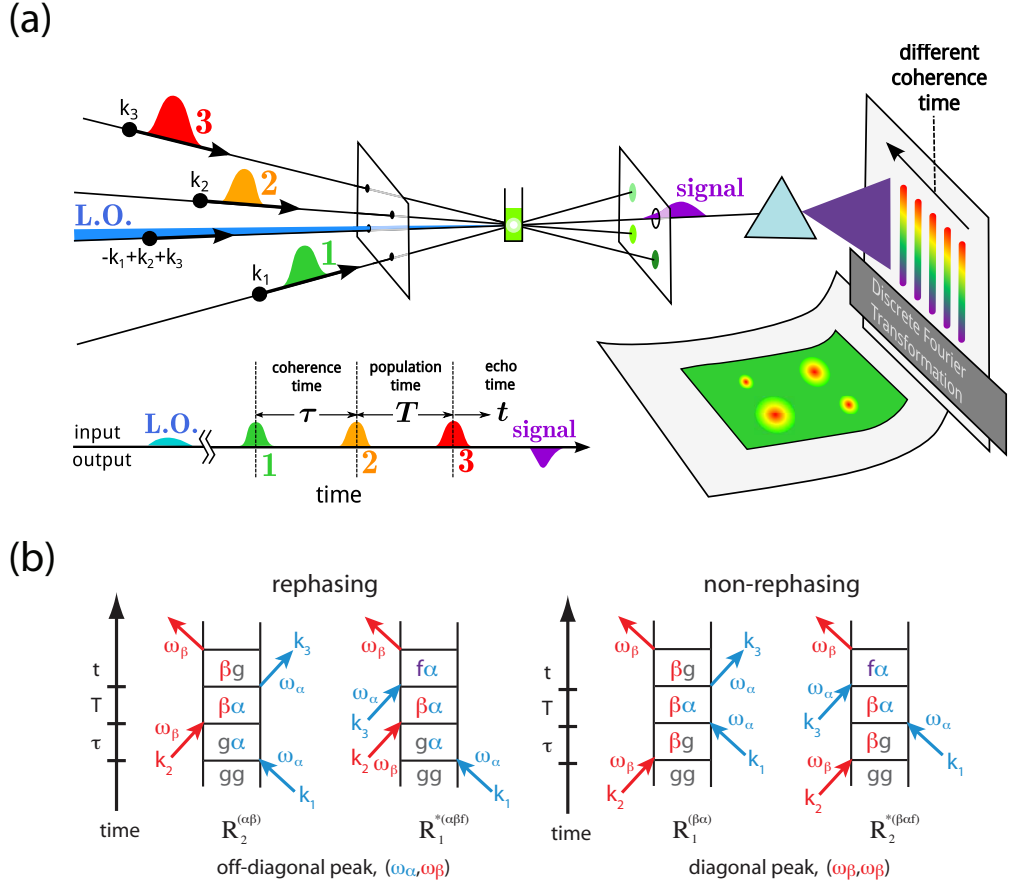}
\caption{\label{coherence_pathways} (Color online) (a) Schematic
of a
  2DES setup. (b) Liouville pathways
  that contribute to the beating signals in the 2DES
  spectra. Contributions to 2DES signals are often depicted
  as double-sided Feynman diagrams as shown here. These diagrams represent the
  pulse-induced time evolution of the density matrix of the system. Light-matter
  interactions are depicted as arrows. An incoming arrow indicates an
  absorption by the system, whereas an outgoing arrow represents a
  stimulated emission event. Density-matrix elements are depicted in
  the center. For example, in the $\mathbf{R}_2^{\alpha\beta}$
  pathway, the first pulse creates a coherence (superposition) between
the ground state $|g\rangle$ and the excited-state $|\alpha\rangle$,
the second pulse then creates a coherence between $|\alpha\rangle$ and
$|\beta\rangle$, and the third pulse prepares a coherence between
$|g\rangle$ and $|\beta\rangle$, which then emits the signal. The
pathway will record a cross-peak at
$(\omega_\alpha,\omega_\beta)$. Because during the population-time
period $(T)$ the system is in a excitonic superposition state, the
signal will carry a phase factor that depends on $T$, leading to quantum
beats on the 2D spectra. The rephasing and non-rephasing contributions
are different in the time ordering of the first two pulses.}
\end{figure}

As mentioned, the advent of powerful experimental tools have
opened-up the ability to probe quantum effects in photosynthetic
systems with unprecedented sensitivity. In particular, 2D
electronic spectroscopy provides a unique tool that is
specifically sensitive to electronic coherence in excitonic
systems
\cite{Abramavicius:2009p75282,cheng2009dynamics,Ginsberg:2009p76649,SchlauCohen:2011cw}.
A 2D experiment is a four-wave mixing process, in which three
laser pulses interact with the sample to create an electronic
polarization that generates the signal (Fig.
\ref{coherence_pathways}a). In the experiment, two periods of time
delay can be controlled in the apparatus: the time delay between
the first and second pulses ($\tau$, coherence time) and time
delay between the second and third pulses ($T$, population time).
At fixed $\tau$ and $T$, the signal field emitted in the
photon-echo phase-matching direction $k_s = -k_1 + k_2 +k_3$ is
combined with an attenuated local-oscillator pulse for
heterodyne-detection and frequency resolved to obtain a spectrum,
which can be regarded as the Fourier transform of the oscillating
signal fields with respect to the time delay between the third
pulse and the signal ($t$, re-phasing time). The measurement can
be repeated at varied $\tau$ and Fourier transformed with respect
to $\tau$ to obtain a 2D spectrum in frequency domain
($\omega_\tau$ and $\omega_t$) at a fixed population time
$T$~\cite{Ginsberg:2009p76649,Abramavicius:2009p75282}.

Because 2D electronic spectroscopy records the signal at the level of the
field rather than the intensity, it is sensitive to the quantum
phase evolution of the electronic system during the population time. A broadband pulse can
then interact with multiple exciton states to produce
superpositions (coherences) of them, and the induced coherence in
the exciton basis then undergoes oscillatory phase evolution that
leads to beating signals as a function of time
$T$~\cite{Cheng:2008p51139}.
Figure \ref{coherence_pathways}b shows the Liouville pathways, i.e.
pulse-induced density-matrix dynamics of the system, that
contribute to the beating signals.  The time evolution of the
coherence during the population time $T$ has an oscillating phase
factor, resulting in quantum beats in the 2D spectra as a function
of $T$. The signal provides an incisive tool to probe quantumness
of an electronic process.

\begin{figure}[]
\begin{center}
\includegraphics[width=\columnwidth]{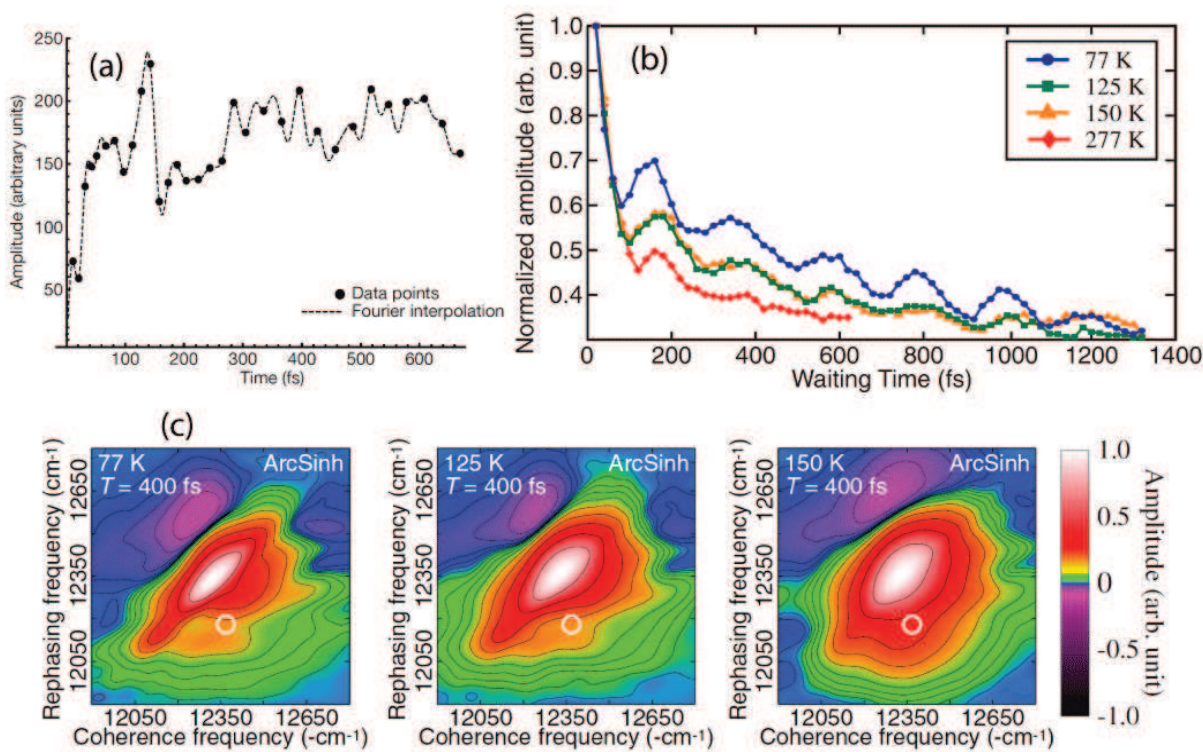}
\end{center}
\caption{{} (Color online) (a) The first example of quantum
coherent beating between excitons 1 and 3 in the FMO complex at
liquid-Nitrogen temperature, from
Ref.~\cite{EngelNature07}. (b) In another experiment (Ref.~\cite{PNAS}), signals of quantum beating
demonstrate the temperature dependence of
 the coherence between excitons 1 and 3, from cryogenic up to ambient room
temperature.  Note that by inspection it appears that coherent
beating is observable in the room temperature data for
approximately 300 fs. (c) An example of the 2D electronic spectra
of the FMO used to collate figure (b). The data for figure (b) is
extracted from the point indicated by the white circle, which is
expected to contain the information about coherence between excitons
1 and 3. The two-dimensional electronic spectra data is collected
by applying a series of three pulses to the sample, and observing
the output signal. Theoretical models indicate that the cross-peak
(which an off-diagonal peak in terms of the whole spectral
picture) in the signal spectra directly corresponds to the
presence of coherence between different excitonic sites.
Mathematically this is equivalent to a kind of quantum process
tomography \cite{Guzik11,Guzik112}, or a probing of the memory of
the system. \label{coherence}}
\end{figure}

As we have described, the FMO complex~\cite{Fenna75} is only one component of many types of light-harvesting
 complexes found in nature, but it is perhaps one of the
most well studied and characterized.
In the seminal work of Engel~\textit{%
et al.}~\cite{EngelNature07}, 2D electronic spectra of
the FMO BChl complex were obtained at $77$K [see
Fig.~\ref{coherence}(a)], and strong quantum beating in the
amplitudes and shapes of the diagonal peaks were revealed.  These
results have been interpreted ~\cite%
{Plenio08,Lloyd08,Caruso09,cheng2009dynamics,LeeCheng09} as a
strong signature of coherent wave-like transfer
of excitation energy in the complex.  More recently, experimental
evidence has shown that quantum coherence survives in the FMO
complex at physiological (ambient room) temperatures for at least
300 fs \cite{PNAS}.  While much shorter than the coherence
observed at cryogenic temperatures [see Fig.~\ref{coherence}(b)],
it is still thought to be sufficiently long to have an impact on
the efficiency of the transport process (which, overall, is of the
order of $1000$ fs). This result is supported by a variety of
theoretical models \cite{Akihito09} which we will summarize
shortly. Similarly, in the experiment performed by Collini
\textit{et al}.~\cite{ambientTemperature}, two light-harvesting
proteins, isolated from marine cryptophyte algae, were found to
have relatively long-lasting excitation oscillations at room
temperatures. This is a particularly interesting example, as the
algae live in exceedingly low-light conditions, and must process
the excitation energy transfer extremely efficiently to survive.
In addition, the coupling of the LHC systems to their protein
environments differs in this case from most other photosynthetic
LHCs, and Collini \textit{et al.} \cite{ambientTemperature}
speculate that this may enhance correlated motions of the protein
environment, enhancing any ``coherence''-preserving effect that
may arise. 

Because short pulses are used in 2D experiments, it is possible
that the quantum beating signals in 2D spectroscopy are results of
vibration-coherence instead of electronic-coherence
\cite{Mancal:2010dj,Nemeth:2010iz}. In principle, this alternative
interpretation can be verified by examining the rephasing and
non-rephasing spectra separately because vibrational-coherence
would cause beating signals in all the rephasing and non-rephasing
signals, whereas electronic-coherence would cause beating in the
off-diagonal peaks of the rephasing signals or the diagonal peaks
of the non-rephasing signals \cite{Cheng:2008p51139}. Recently,
Turner and coworkers \cite{Turner:2011p99265} have carefully
examined vibrational coherence effects and electronic effects in
2D spectroscopy and confirmed that quantum beating observed in the
PC645 light-harvesting antenna protein of the cryptophyte alga
{\it Chroomonas sp.} is due to electronic coherence.

In addition to 2D spectroscopy, other nonlinear optical
experiments can be made into sensitive tools for quantum coherence
phenomena. By using a two-color photon-echo approach, Lee
\textit{et al.}~\cite{LeeScience07} also reveals that electronic
coherence between bacteriopheophytin (H) and accessory
bacteriochlorophyll (B) in the reaction center from the purple
bacterium {\em R.~sphaeroides} is preserved for much longer time
than would be expected from the dephasing of either chromophores
both at 77K and 180K \cite{LeeScience07}. Theoretical modeling
revealed that the long-lasting coherence is indicative that the
fluctuations of transition energies of B and H are strongly
correlated in the reaction center. It was speculated that the
``coherence preservation effects'' from correlated protein
environments also explain the surprisingly long coherence time (of
the same order as the time for the excitation to traverse the
whole complex) observed in FMO and other light-harvesting
complexes.

\subsection{Theoretical models}

To gain a better understanding of what these various results imply
it is helpful to consider a model commonly used to simulate some
of these systems. One approach to simulate the quantum behavior of
a single excitation in the FMO complex is via an effective Frenkel exciton
model \cite{vanAmerongen:2000tb,Blankenship}.  As described earlier, a
photon creates an excitation in an antenna molecule, which is
transferred to one of the sites in the FMO molecule, which is then
transferred along the chain, from chromophore to chromophore via a
transition-dipole Coulomb interaction, until reaching the reaction center,
where it
is then employed for charge separation.
The assumption of a single-excitation seems valid for {\em in
vivo} situations, particularly when the bacteria is living in very
low-light conditions. It is exceedingly unlikely for the FMO
complex to ever contain two excitations. Although the
single-excitation assumption in experimental situations has
generated certain controversy ~\cite{Tiersch11}, rapid
exciton-exciton annihilation processes makes multiple-exciton
states in a single complex extremely unstable and tend to relax
into a single-exciton state before energy transfer occurs.
Nevertheless, in principle it is straightforward to construct a
multiple-excitation model, e.g., as employed in \cite{Ghosh} to
describe an artificial photosynthetic system.  For the purposes of
clarity, however, we restrict ourselves here to the
single-excitation model, which can be written as
\begin{equation}
H=\sum_{j=1}^{N}\epsilon _{j}|j\rangle \langle
j|+\sum_{j<j'}J_{j,j'}(|j\rangle \langle j'|+|j'\rangle \langle
j|),\label{Hamiltonian}
\end{equation}%
where the states $|j\rangle $ represent the presence  of an
excited electron (exciton) at site, or BChl molecule, $j$, where
$j\in 1,..,7$ (see Fig. 1(b) for a figurative description of these
sites in the FMO complex, and we omit the eighth BChl molecule because it was
not discovered in structural models of the FMO complexes until recently). In
other words, one can describe the whole FMO complex with a single
label defining at which site the excitation is residing. The
parameter $\epsilon _{j}$ is the energy of that excitation for a
particular site (which is dependent on the surrounding protein
structure and varies quite a lot), and $J_{j,j'}$ is the excitonic
coupling between the $j$ and $j'$ sites.

The determination of the various energies and coupling strengths
in a given LHC is a complicated and difficult area of research,
often involving both spectroscopy and {\em ab initio} modeling of the
physical structure and environment of the system
\cite{cheng2009dynamics,Renger:2009p77537}. In general, accurate values of electronic
couplings that are in good agreements with experiments can be
calculated with  {\em ab initio} quantum chemistry methods based on
the atomistic models of a LHC
\cite{Hsu:2009p70929,Renger:2009p77537,Mennucci:2011p98662}. The site
energies are more difficult to determine because modeling the
protein-pigment interactions is a non-trivial task, therefore
experimental inputs are usually required in order to obtain accurate
energies. Recently, Renger and coworkers have demonstrated in several
photosynthetic
complexes \cite{Adolphs:2008p27712,Mueh:2010p89233,Adolphs:2010p80273}
that with a careful treatment of the electrostatic interactions
between the electronic
excitations and the surrounding protein environments, it is possible
to calculate site-energy parameters with structure-based theoretical
approaches.

Since the FMO LHC is one of the most studied examples, some
understanding of the system's Hamiltonian exists (though often
different values for the energies and excitonic coupling
amplitudes are given in the literature \cite{Olbrich11-2,
Ritschel,Adolphs:2008p27712}). We now present the values used in
Ref.~\cite{Adolphs06},
\begin{eqnarray} \label{Hamiltonian2}
H= \left(%
\begin{array}{ccccccc}
215 & -104.1 & 5.1 & -4.3 & 4.7 & -15.1 & -7.8 \\
-104.1 & 220 & 32.6 & 7.1 & 5.4 & 8.3 & 0.8 \\
5.1 & 32.6 & 0 & -46.8 & 1.0 & -8.1 & 5.1 \\
-4.3 & 7.1 & -46.8 & 125 & -70.7 & -14.7 & -61.5 \\
4.7 & 5.4 & 1.0 & -70.7 & 450 & 89.7 & -2.5 \\
-15.1 & 8.3 & -8.1 & -14.7 & 89.7 & 330 & 32.7 \\
-7.8 & 0.8 & 5.1 & -61.5 & -2.5 & 32.7 & 280%
\end{array}%
\right)
\end{eqnarray}
where units are cm$^{-1}$ and the large energy gap relative to the
common ground state has been subtracted.   The diagonal elements
correspond to the energies $\epsilon_j$ in Eq.
(\ref{Hamiltonian}), while the off-diagonals correspond to the
site-site electronic couplings $J_{j,j'}$.  We present these numbers here to
show that the site-site couplings are of the same order as the
energy difference between sites.  Already this suggests that
site-site coherences could be strong.  In addition, the magnitudes of
site-site couplings are often distributed broadly, with the larger
ones in the scale of $\sim$100 cm$^{-1}$, which is also comparable to
the reorganization energy due to exciton-environment couplings.

\subsubsection{Approximations used in the models}

One of the open questions for this system is the nature of its interaction with its
environment.  In the simplest model, which we explicitly show below, a variety
of approximations are made. The first is to assume that each site
is coupled to a bath of oscillators, and that the
baths coupled to each site are independent of each other. The second
approximation is to assume that this coupling is to the
energy of each site~\cite{Cheng07, Plenio08,Palmieri,pollard96,
fleming02}, which means that the
environment causes fluctuations of the energy in the site basis.
In the exciton basis, the fluctuations become to have off-diagonal
matrix elements that can cause transitions between
eigenstates \cite{Lloyd08}.
The third is to assume the environment is Markovian
(i.e., without memory). Finally, the fourth assumption is to assume the
exciton-bath coupling is weak, and that one can perform a second-order
perturbation theory to describe the dynamics of energy transfer.

This gives a master equation that describes the reduced density
matrix of the FMO complex that is simple to solve, and has a
minimal number of free parameters. Typically there is also a
temperature-independent radiative relaxation rate for each site.
If the excitation takes too long to traverse its way through the
complex, it will be eventually lost due to this relaxation.
Fortunately, fluorescence decay of chlorophylls typically occurs
in the nanosecond time scale, a slow process compared to the other
dynamics, ultimately giving the FMO complex its ability to
efficiently transport energy with a high success rate.  This
simple model has estimated some of the qualitative properties of
experiments on FMO, including the long coherence time, and high
transport efficiency \cite{Caruso09}.  

As a specific example, the Markovian master equation approach
 assumes a self-energy,
\begin{equation}
\Sigma[\rho]=\Sigma_{\text{ss}}[\rho]+\Sigma_{\text{deph}}[\rho]+\Sigma_{\text{diss}}[\rho],\label{seng}
\end{equation}
where $\rho$ denotes the system density matrix. The first term
describes irreversible excitation transport between site $3$ and
the sink \cite{Adolphs06, Wen09} (see Fig.~1(a) and (b) for a
description of where site $3$ and the sink, or reaction center,
are in the FMO):
\begin{equation}
\Sigma_{\text{ss}}[\rho]=-\Gamma_{D}[s_{D}s_{D}^{\dag}\rho-2s_{D}^{\dag}\rho
s_{D}+\rho s_{D}s_{D}^{\dag}],
\end{equation}
where $s_{D}=\left|3\right\rangle\!\!\left\langle 0\right|$, and
$\Gamma_{D}$ is the sink tunneling rate. Site $3$ is coupled to
the reaction center because it comprises the lowest energy
excitation and lies closest to the reaction center. In contrast,
site $1$ lies closest to the antenna.  Here $\left|0\right\rangle$
denotes the empty state with no electronic excitation. The second
term, $\Sigma_{\text{deph}}[\rho]$, describes a
temperature-dependent dephasing (see Ref.~\cite{Robenstrost} for a
detailed discussion of the temperature dependence).  The last
term, $\Sigma_{\text{diss}}[\rho]$ describes the slow fluorescence
relaxation process.  A variety of insights
have been obtained from this simple model~\cite{Cheng07,
Plenio08} which will discuss in the next section, but the essence is that the combination of coherent dynamics
from $H$ and level-broadening from $\Sigma_{\text{deph}}[\rho]$
means there is a high possibility of bringing an excitation from
site $1$ to site $3$, and thus the RC, before the excitation is
lost due to the fluorescence relaxation
$\Sigma_{\text{diss}}[\rho]$.  The efficiency of the transport is
often modelled via the time-dependent sink or reaction-center
population \beq P_{\mathrm{RC}}(t) = 2\Gamma_{D} \int_0^t
Q^{(3)}(t')~dt',\eeq where
$Q^{(3)}=\textrm{Tr}[\ket{3}\bra{3}\rho(t)]$ is the population of
the site ``$3$",  connected to the sink.

\begin{figure}[]
\begin{center}
\includegraphics[width=0.5\columnwidth]{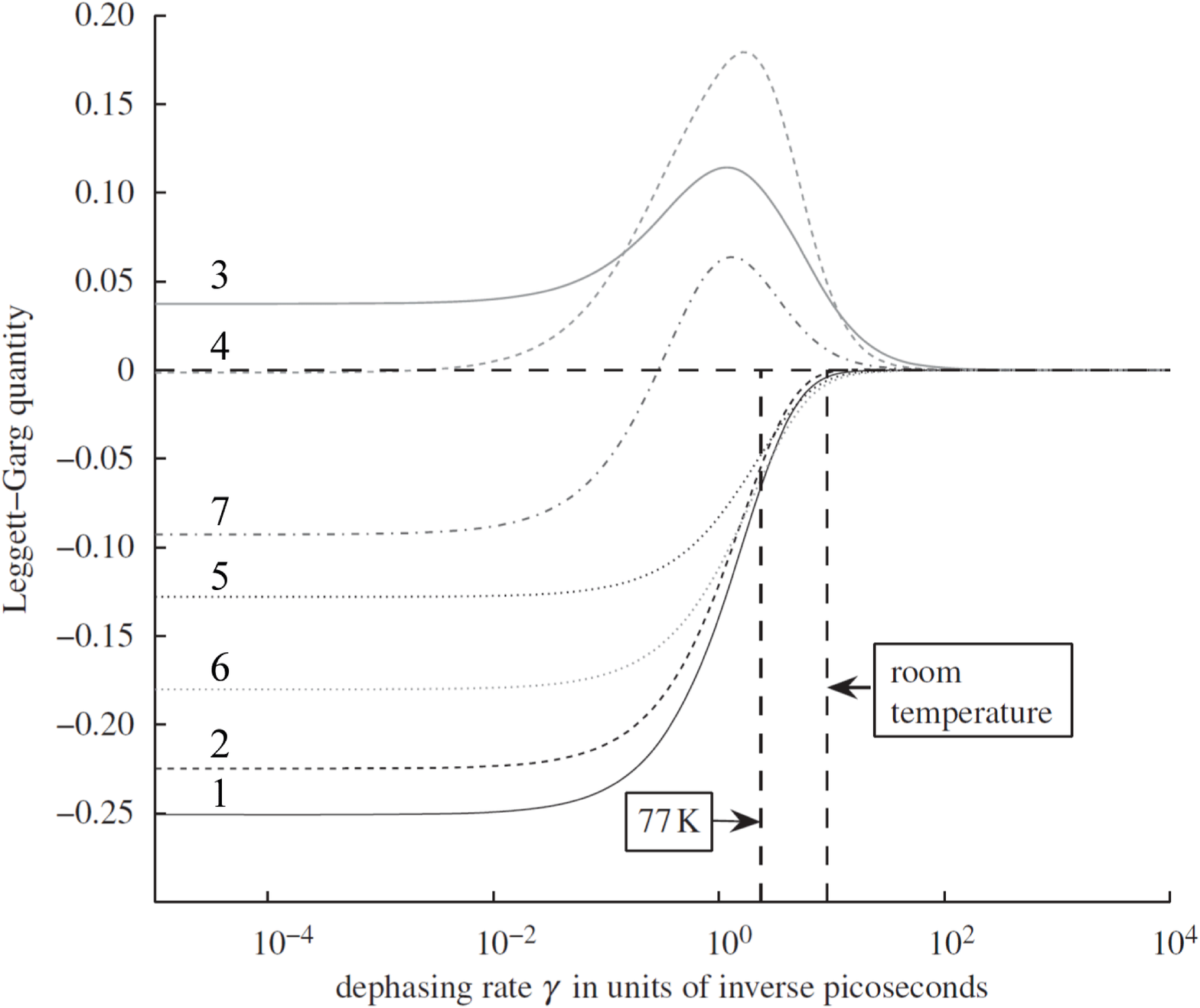}
\end{center}
\caption{{} (Color online) This figure, from Wilde et al
\cite{Wilde10}, shows a theoretical violation of the Leggett-Garg
inequality as a function of dephasing (and hence bath temperature)
for measurements on each site, $1$ to $7$.  A value of $<0$
indicates a violation, and the results suggest quantum phenomena
persists up-to room temperature for measurements on sites $1$,
$2$, $5$, and $6$.  As yet there is no way to directly measure
such non-invasive observables in FMO.  It is particularly
difficult, as unambiguous violations of the Leggett-Garg
inequality in this case would require non-invasive measurements of
excitonic populations. Most fluorescence and spectroscopic
measurements are destructive, and thus not appropriate for
observing the Leggett-Garg inequality.  However, ultimately this
kind of test would help rule out alternative equivalent
explanations for efficient transport in FMO.\label{LG}}
\end{figure}

As mentioned before, in most treatments of the FMO system, the
coupling to the environment modulates the energy of each site, and
in the language of quantum mechanics this is ``pure-dephasing''.
The radiative relaxation of each site is independent of
temperature because the optical transition energy of each site is
exceptionally high ($> 12,000$ cm$^{-1}$), leaving only the
dephasing processes to be temperature dependent.

In the FMO complex, each state in the above description represents
an exciton on a molecule which transfers its energy to its
neighbor, but no actual electron transport takes place. This
implies that the re-organisation energy (the energy associated
with a change in the surrounding protein environment due to the
presence of an electron charge) is relatively low compared to
processes where electron transport occurs. In general, however, both the assumption of independent Markovian
baths and of weak coupling to these baths may not be realistic.
There is evidence that the coupling to protein environments around
the FMO complex can be of the same order as the electronic
couplings~\cite{Robenstrost,Ghosh} ($100$ cm$^{-1}$).  In
addition, the bath (e.g., the protein scaffold surrounding the FMO
complex) may have structure and dynamics which have a strong
correlation with, and back-action on, the dynamics of the
excitation in the FMO complex. To understand the effect of this
intermediate coupling regime, and complex environment, the system
and bath have been treated with a range of non-Markovian and
higher-order (in electron-phonon coupling) models
\cite{Akihito09, Cheng07, Jang:2011ee,Jang:2008p53203, Ishizaki10, Peter11, Ritschel,Ghosh,Kolli:2011ki}.  
In particular, nonperturbative approaches that provide exact
numerical results for certain models of excitation energy transfer
have been applied to study coherent quantum dynamics in
photosynthetic light harvesting
\cite{Ishizaki:2009p75287,Jin:2008p60247,Nalbach:2011p100197,Chen:2011bm}.  We
will discuss the details of one of these approaches (the hierarchy
model) later. Although these nonperturbative methods are
computationally too expensive to apply to large photosynthetic
complexes, they have provided valuable insight that have shed a
different light on the coherence effects in light harvesting in a
different light \cite{Akihito09}; one where bath memory effects
conspire to enhance quantum coherence.

Going beyond the independent bath approximation, initial
experiments indicated that the surprising long-lasting quantum coherence
between two electronic states may be because of coherence
enhancing effects from coupling to common vibrational modes
\cite{Cheng07, LeeScience07,ambientTemperature}. This has
triggered the reexamination of the excitation transfer processes
\cite{Nazir09, Silbey09} and decoherence effects \cite{Hoda10,
Fleming11} in the presence of such an unusual environment.  In
principle, such a strongly coupled ``common bath'' should contain
bath-induced fluctuations that are spatially correlated
~\cite{Cao1,LHIICorrellated}. As a counter-argument, however, some
molecular dynamics simulations for the FMO complex~\cite%
{Olbrich11}  and reaction center~\cite{Jing:2012kb} show that only
weak correlations between the movements (vibrations) of the chromophores appear, and that the uncorrelated bath approximations {\em
may} be valid.

Very recent work by Shim \textit{et al} \cite{Shim11} compared
some of these master-equation models to a more complex many-body
atomistic/molecular modeling scheme and found excellent agreement
between the two approaches. This is encouraging as it does suggest
that the assumption (based on relatively simple models) that the
oscillations seen in experiment have a truly quantum origin might
be correct. However, whether the environment surrounding LHC
systems has a structure or nature that preserves quantum effects
in some way is still a controversial and open question. More
experimental work is required to answer this puzzle.

Finally, some authors have argued that it is possible to construct
a variety of alternative classical models that can, in principle,
produce both classical beating \cite{Briggs11} and efficient
energy transport \cite{Cheng07, Plenio08} without reference to
quantum coherence at all.  Such alternative descriptions must be
eliminated before one can unambiguously state that the FMO
complex, or other LHCs, take advantage of quantum mechanics.


\subsection{New Insights from Theoretical Modelings}

\subsubsection{Environment-assisted transport}

Apart from the direct observation in experiments of quantum
coherent beating, some of the most intriguing insights that have
appeared from the study of the FMO complex suggest that a
combination of these quantum coherent oscillations of excitations
between sites, and interactions with the environment, produce a
transport efficiency higher than is possible with a F\"{o}rster
model \cite{Forster48} alone.  If correct, this would fit the
definition of functional quantum biology. Such a phenomenon was
proposed and studied by Plenio \textit{et al} \cite{Plenio08}, by
Mohseni \textit{et al} \cite{Lloyd08},  and by Lee \textit{et al}
\cite{LeeCheng09}. In all cases, it is assumed that a single
excitation is placed at a particular site in the FMO (usually site
one), and that this excitation then propagates through the FMO
chain due to a combination of coherent tunnelling and
environmental effects. As mentioned earlier, the goal of the FMO
complex is to get this excitation efficiently to site three, which
is coupled (incoherently) to a reaction center. Thus the typical
time, or efficiency, of reaching the reaction center is calculated
as a function of environment temperature, site
selective-couplings, and so on. In most cases it seems that the
full quantum model (e.g., based on Eqs. [1-3]) will give a
 higher efficiency than a purely classical one. Several
physically intuitive explanations have been given for this
phenomenon:

\begin{itemize}
\item Local minima avoidance:  Refs.~\cite{Plenio08} and
\cite{Lloyd08} showed that the Markovian quantum model outlined
earlier suggests both long-lived coherent oscillations between
sites, and an enhanced rate of excitation transfer to reach the
``sink'' site, over that predicted by the F\"{o}rster model.
Reference \cite{Plenio08} argued that essentially the combination
of coherent transfer and environmental ``noise'' causes level
broadening, which implies that the excitation can more easily
escape local minima in the FMO network (see Fig. [1]).  In other
words, quantum delocalization can help avoid and overcome local
minima, or energy potential traps, in the energy landscape of the
FMO complex \cite{Akihito09}.  It has been suggested that such a
phenomena is even more important in higher plants because of
possible ``up-hill'' potential landscapes.  These results are
related to earlier work by Gaab and Bardeen \cite{Gaab} who
outlined how energy transfer through a network can be optimized by
carefully choosing the coherence time and the rate of transfer to
the sink.

\item Coherence assisted trapping:  In a study by Lee \textit{et
al.} \cite{LeeCheng09}, it was demonstrated that coherent quantum
dynamics, together with rapid incoherent dissipation due to a
trapping site can enhance the efficiency of the irreversible energy transfer between
the initial energy donor and the sink, or reaction center, over a
purely classical model. Effectively, this model suggests that an
intricate combination of reversible quantum coherent evolution and
incoherent collapsing of exciton wave functions can achieve a
greatly enhanced energy-trapping efficiency, effectively using the
anti-Zeno effect to promote energy transfer.

\end{itemize}

However, one should again notice that this ``enhanced efficiency''
is in comparison to that obtained with the F\"{o}rster model.
There remain several ambiguities; Firstly, is the F\"{o}rster
model the correct classical model to make a comparison to?
Secondly, placing a single excitation at site ``$1$'' may match
well certain experiments, but in normal light conditions is this a
correct approach?

Quite recently, Briggs and Eisfeld \cite{Briggs11} showed that,
for realistic coupling strengths, an alternative classical model
of the FMO transport process can produce results identical to the
quantum one.  In addition, a recent analysis by Wu \textit{et al.}
\cite{Cao2} suggests that, as mentioned earlier, the efficiency
enhancement gained by coherence in the FMO complex is only a few
percent compared even to that predicted by the F\"{o}rster model.
Finally, the true {\em in-vivo} conditions are not well understood,
and is the subject of current research.
Perhaps the only way to really solve this issue is to show that in
more complex components of some light harvesting complex, e.g., in
LH1 or LH2 \cite{LHII, LHIICorrellated}, there exist significant
energy traps which, without the assistance of quantum coherence,
drastically impact/reduce the probability of an excitation
successfully navigating its way to a reaction center before being
lost to fluorescence relaxation.

\subsubsection{The Hierarchy model}

In an effort to gain a deeper understanding of the interplay
between the quantum coherent transport of energy in FMO and how it
interacts with its complex protein environment, a variety of
non-Markovian and non-perturbative models have been applied to
these systems. One of the most successful is the Hierarchy model,
originally developed by Tanimura and Kubo~\cite{Tanimura3}, which
has been applied to both the FMO complex~\cite{Akihito09} and
other LHC components~\cite{LHII, LHIICorrellated}.  This model has
a had a large impact this field, and thus we will briefly describe
its main components here. Full derivations can be found in the
literature \cite{Tanimura, Tanimura2, Tanimura3,
StochasticMethod}. Starting with the Hamiltonian we discussed
earlier, one explicitly describes the interaction term between
site energies and phonon modes, \beq H^{(e-p)} = \sum_j
\ket{j}\bra{j} \; \hat{q}_j\eeq where $\hat{q}_j = -\sum_k g_{j,k}
x_k$, and $g_{j,k}$ is the coupling constant of the $j$th site and
the $k$th mode in the bath.  The phonon modes themselves are
described as harmonic oscillators (with a Hamiltonian $H^{(p)}$
which we omit here for brevity).

It is typically assumed that at $t=0$ the sites (or pigments) and
the phonon modes are separable, so that $\rho(0)=\rho_e(0)\otimes
\rho_p(0)$, and the phonon modes are in a thermal equilibrium
state $\rho_p(0) = e^{-\beta H^{(p)}}/\mathrm{Tr}\left[e^{-\beta
H^{(p)}}\right]$, $\beta = 1/K_{\mathrm{B}}T$.  After averaging,
the correlation function of the bath modes \beq C_j(t) =
\ex{q_j(t)q_j(0)} = \frac{1}{\pi} \int_0^{\infty} \!\!\! d\omega
\; J_n(\omega) \frac{e^{-i \omega t}}{1-e^{-\beta\hbar
\omega}}\eeq are sufficient to describe the properties of the
bath. With the Hierarchy method one typically uses a Drude
spectral density (appropriate for over-damped oscillators) \beq
J_j(\omega) = \left(\frac{2\lambda_j \gamma_j}{\hbar}\right)
\frac{\omega}{\omega^2 +\gamma_j^2}. \eeq Here $\gamma_j$ is the
``Drude decay constant'', and indicates the memory time of the
bath for site $j$ (each site is assumed to have its own
independent bath, though in general one of the powers of the
Hierarchy method is its ability to treat correlated baths
\cite{LHIICorrellated}).  Also, $\lambda_j$ is the reorganisation
energy, related to the system-bath coupling strength.  This Drude
spectral density implies an exponentially-decaying correlation
function, \beq C_j = \sum_{m=0}^{\infty} c_{j,m}
\exp\left(-\mu_{j,m} t\right) \eeq where $\mu_{j,0} = \gamma_j$,
$\mu_{j,m}\geq 1 = 2\pi m/\hbar \beta$, and the coefficients \beq
c_{j,0} = \gamma_j \lambda_j\left(\cot(\beta \hbar \gamma_j/2) -
i\right)/\hbar\eeq and \beq c_{j,m\geq 1} = \frac{4\lambda_j
\gamma_j}{\beta \hbar^2} \frac{\mu_{j,m}}{\mu_{j,m}^2
-\gamma_j^2}. \eeq

\begin{figure}[]
\includegraphics[width=\columnwidth]{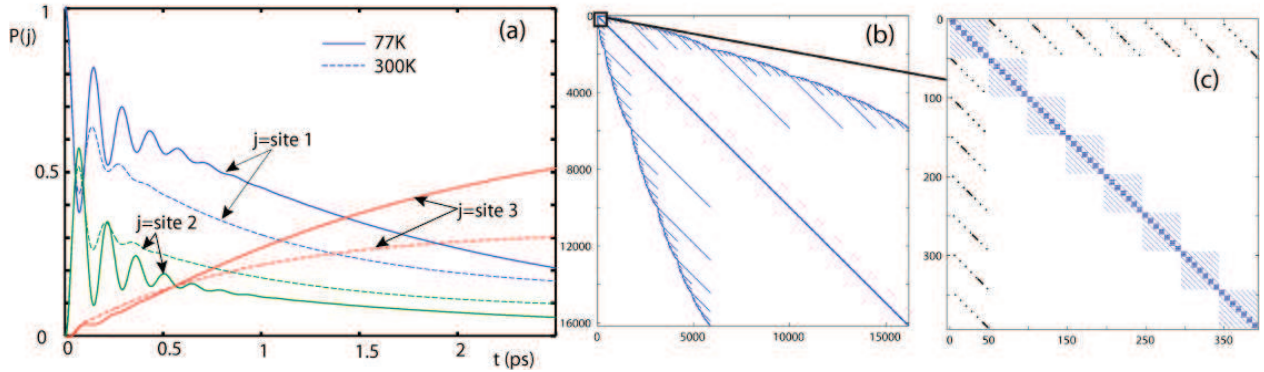}
\caption{{}\label{Hierarchy} (Color online) (a) The population
dynamics predicted for the FMO complex by the hierarchy model.
The parameters used were $\gamma^{-1} = 50$ fs and $\lambda=35$
cm$^-1$, for all sites, as in \cite{Akihito09}. The dashed lines
indicate a bath temperature of $300$ Kelvin and solid lines $77$
Kelvin. Truncation was taken at $K=0$ and $N_c=4$.
 The lower figure is a representation of the sparse matrix of
coefficients describing the coupled
 hierarchy equations.  For $N_c=4$ and $N=7$, there are
$(N_c+N)!/(N_c! N!)= 330$ total density operators in the
hierarchy.  Each density operator obeys the dynamics of
 the $7$-site model, and thus in a typical open-quantum-system
super-operator basis is described by $49$ equations.  In total
this gives a matrix of equations of size $16,170$.
 It is helpful to present this matrix in (b) and (c) as it serves as a visual
check for faults in implementation (e.g., due to missing symmetry
or so forth).  The magnification of the $8\times 49=392$ elements,
in (c), illustrates
 how the system density matrix is coupled to the seven $N_c=1$ tier
of auxiliary density matrices (the coupling terms are highlighted
in black).}
\end{figure}

With just this information on the bath properties (the
reorganisaiton energy, bath memory time, and bath temperature) one
can employ the Hierarchy equations to describe the system-bath
dynamics in the strong-coupling and non-Markovian regime, \beq
\dot{\rho}_{\mathbf{n}} &=& -(i L + \sum_{j=1}^N\sum_{m=0}^K
\mathbf{n}_{j,m} \mu_m) \rho_{\mathbf{n}} -
i\sum_{j=1}^N\sum_{m=0}^K\left[Q_j,\rho_{\mathbf{n}_{j,m}^+}\right]\nonumber\\
&-& i\sum_{j=1}^N\sum_{m=0}^K
n_{j,m}\left(c_mQ_j\rho_{\mathbf{n}_{j,m}^-} - c_m^*
\rho_{\mathbf{n}_{j,m}^-}Q_j\right). \eeq Here,
$Q_j=\ket{j}\bra{j}$ is the projector on the site $j$, $L$ is the
Liouvillian described by the Hamiltonian introduced earlier, and
for FMO $N=7$. The various parameters are those we defined above.
The Hierarchy is a large set of coupled equations each labelled by
$\mathbf{n}$, a set of non-negative integers uniquely specifying
each equation. The integers are defined as
$\mathbf{n}=\{n_1,n_2,n_3,...,n_N\} =
\{\{n_{10},n_{11},..,n_{1K}\},..,\{n_{N0},n_{N1},..,n_{NK}\}\}$.
That is, each site $j$ has an additional label $m$, from $0$ to
$K$, and each of those labels in turn can run from $0$ to
$\infty$. The label $\mathbf{n}=0=\{\{0,0,0....\}\}$ is special,
and refers to the system density matrix.  Its properties at any
time $t$ define those of the system. This is in turn coupled to so
called ``auxiliary density matrices'', which describe the complex
bath fluctuations, by the terms in the equation where
$\mathbf{n}_{j,m}^{\pm}$ (i.e. $\mathbf{n}_{j,m}^{\pm}$ implies
the term in the index defined by $j$ and $m$ is increased or
decreased by $1$). As a simple example, if the terms in the sum
for $m$ are truncated at $K=0$, then the system density matrix is
coupled to $N$ other auxiliary density matrices directly,
$\mathbf{0_1^+}=\{1,0,0,...0\}$, $\mathbf{0_2^+}=\{0,1,0,...0\}$,
and so on (see Fig.~\ref{Hierarchy}(c)).  Each of these auxiliary
equations is then in turn coupled to higher-tier equations, and
overall the hierarchy forms a high dimensional simplex.

The hierarchy equations are infinite both in the sum of the $m$
label (due to the expansion of the correlation function), and in
the value of each label itself. Thus both must be truncated in
some way. Typically, the $m$ label can be truncated if $\beta
\hbar \gamma_m < 1$ and $2\pi (K+1)/\hbar\beta > \omega_S $ (where
$\omega_S$ is some characteristic system frequency).   The overall
truncation of the labels is defined by the largest total number of
terms in a label, or the tier, $N_c = \sum_{j,m} n_{j,m}$.  This
concept is important as each ADO in a given tier is only coupled
to those ADOS in the tier above or below it.  In Yan \textit{et
al} \cite{StochasticMethod} they describe this structure, and the
meaning of each tier, as the level of bath self-correlation
included in the simulation.

The tier at which the truncation should be made is difficult to
predict (there is no small parameter here), though typically
truncation should be taken when $N_c \gamma
>> \omega_S$ (though it also depends on the reorganisation energy and temperature).
In practice, truncation should be checked by looking for
convergence in system dynamics when changing $N_c$. A systematic
and powerful algorithm to find convergence was introduced by Shi
\textit{et al} \cite{Shi09}. They proposed that a renormalization
of the hierarchy equations allows a direct inspection of the
elements of higher auxiliary equations to be made during the
numerical solution, and thus the level of the tier can be
truncated as necessary.

In addition, a temperature correction can be added to the
equations of motion, which compensates for the truncation of the
$m$ label at level $K$, and assists in overall convergence.  This
is made by assuming that all higher-level auxiliary density
operators for $m>K$ are uncoupled from each other and that they
experience pure Markovian dynamics \cite{Tanimura}. This results
in an additional term in the equation of motion, sometimes called
the ``Ishizaki-Tanimura boundary condition'' \cite{Tanimura},
which can be included in the Hierarchy equation of motion; \beq
L_{\mathrm{IT-BC}} =- \sum_{j=1}^N\sum_{m=K+1}^{\infty}
\frac{c_{j,m}}{\mu_{j,m}}
\left[Q_j,\left[Q_j,\rho_{\mathbf{n}}\right]\right].\eeq The
double commutator gives a normal Lindblad form, and the summation
can be done analytically, e.g., for $K=0$, \beq
\sum_{m=1}^{\infty} \frac{c_{j,m}}{\mu_{j,m}} =
\frac{4\lambda_j}{\hbar^2 \gamma_j
\beta}\large[1-\gamma_j\hbar\left(\cot(\gamma_j\hbar
\beta//2)\right)\beta/2\large].\eeq  In other words, this term
encapsulates additional environmental dephasing effects neglected
by truncating $K$.  Including this correction, even for $K=0$, has
been found to substantially assist the convergence of the
Hierarchy results. As an example, in Fig.~(\ref{Hierarchy}), we
show the results of an application of this model to the FMO system
at $77$ K and $300$ K, for various $\gamma_j$ and $\lambda_j$
employed in the literature. The beauty of this method is that, to
some degree, all parameters can be extracted from experiments and
the results correctly predict the long coherence time seen in
experiments. Perhaps the most important point here is that long
bath memory (implied by small $\gamma_j$) helps increase the
coherence time of the coherent oscillations \cite{Akihito09},
though this may not necessarily enhance the overall transport
efficiency. In general, for each parameter, there is an optimal
value which gives the largest transport efficiency from site $1$
to site $3$, and hence to the reaction center \cite{Cao1,Cao2}.

\subsubsection{Leggett-Garg inequality and entanglement measures}

As mentioned earlier, it is important to fully verify whether the
experimental observations of coherence truly come from quantum
mechanics and not some alternative, classical, model
\cite{Briggs11}. Wilde \textit{et al} \cite{Wilde10} proposed
using the Leggett-Garg inequality as a means to unequivocally
ascertain whether the dynamics observed in these experiments are
quantum or obey ``macroscopic realism''.  As yet, however, the
measurements needed for such a test cannot be performed in these
FMO experiments.  However, their results show that if it were
possible to perform such measurements, the currently accepted
physical parameters for the FMO system suggest a violation of the
inequality at room temperatures should occur (see Fig.~\ref{LG})
for an example of the violation of the inequality.

In addition to coherence, it is also interesting to ask whether
there is appreciable entanglement in the FMO complex during the
exciton transfer. It was recently suggested by Sarovar \textit{et
al.}~\cite{Sarovar10} that a small amount of long-range and
multipartite entanglement should exist even at physiological
temperatures \cite{Sarovar10}, however, the role of entanglement
in the energy transfer process is still not clear (see also
\cite{Plenio102}).

\subsubsection{Quantum process tomography}

Quantum state tomography is a well-known technique to reconstruct
the full density matrix or state of a system. Recently, there has
been increasing attention on what is termed quantum
\textit{process} tomography \cite{Cirac97, Lidar06}, which
attempts to fully characterize a given quantum process (i.e., a
systems dynamics) in an open environment (e.g., in contact with
Markovian and non-Markovian baths). Simply speaking, quantum
process tomography is a systematic procedure to characterize an
unknown quantum system, or black box, by systematically analyzing
the
functional relationships between inputs and outputs as a function of time. 
Very recent works \cite{Guzik11,Guzik112} showed that the 2D
electronic spectra used to investigate light-harvesting complexes
is equivalent to a kind of quantum process tomography, elucidating
the meaning of the results found with these techniques in earlier
experiments. In principle this means that the spectroscopic
experiments of excitonic systems can be studied from a quantum
information theory approach, and that perhaps the full nature of
the dynamics in FMO and other systems can be understood on a
deeper level. In the future perhaps it may be possible to discuss
important questions such as the existence of non-Markovian
environments, entanglement, and decoherence during the transfer
process using this tool. However, their proposal still requires
steps that are experimentally challenging.

\subsection{Functionality}

One of the arguments used to justify the presence of ``functional
quantum mechanics'' is that if there is any advantage, or
improvement in efficiency, to be had from taking advantage of
quantum mechanics, evolution will inevitably do so. Quantum
coherence, manifested in the delocalized eigenstates of
photoexcitations in photosynthetic complexes, plays fundamental
roles in the spectral properties and energy-transfer dynamics of
photosynthetic light harvesting
\cite{Yen:2011p99898,Scholes:2011kx}. In structures of
strongly-coupled clusters of pigments, coherence is essential for
shaping the energy landscapes and energy flow towards the reaction
center. This has been demonstrated clearly in the LH2 antenna
complex of purple bacteria \cite{Hu02,Jang:2004p1034,Grondelle06},
the FMO complex~\cite{Louwe:1997p29964,Adolphs06}, the reaction
centers~\cite{Jordanides:2001jz}, and the major light-harvesting
complex of higher
plants~\cite{Grondelle06,SchlauCohen:2009p77451}. For example, in
the LH2 antenna complex, energy transfer from a B800 BChl to
coherently delocalized B850 states is an order of magnitude faster
than that to a single B850 BChl
\cite{Jordanides:2001jz,Jang:2004p1034}. Such a significant
enhancement in energy transfer rate clearly contributes to the
highly efficient light-harvesting process, hence the
functionality, in purple bacteria.

A separate issue is whether or not quantum coherent dynamics, i.e.
excitonic coherence, plays a role in the functionality of
photosynthesis. This issue has become the subject of intensive
research recently. However, in reality the calculated gain in
efficiency (as a function of time, for example) seen in
Refs.\cite{Plenio08, Lloyd08, LeeCheng09,Cao1,Cao2} is relatively
small. In some species (e.g., the algae living in low-light
conditions studied by Collini et al~\cite{ambientTemperature})
this small advantage may be useful. But in general, (e.g., plants
in high-level light conditions), is there still a significant
advantage provided by coherence?  As mentioned earlier, some
studies of ``higher plants'' suggest the energy landscape of their
LHCs is even more rugged than that in the FMO
complex~\cite{ambientTemperature}, and thus coherence may become
even more important in efficiently navigating this terrain before
other processes (like fluorescence relaxation) steal away the
excitation. Interestingly, a recent study by Hoyer {\it et al.}
\cite{Hoyer:2010fl} indicates that the effect of ``quantum
speedup'' is rather small in the time scales of photosynthetic
energy transfer, and they propose that quantum coherent effects in
light-harvesting systems are more likely to be optimized for
robustness or total efficiency.

Finally, very recent results by Ritschel et al \cite{Ritschel}
paint an even more complex picture;  they model the full FMO
trimer, including an additional eighth bacteriochlorophyll (BChl)
molecule which was recently found to exist in the FMO complex.  On
a positive note, they find that the subunits of the trimer act as
separate transport channels from the antenna to the reaction
center, thus the assumption of studying the individual FMO
monomers individually seems safe. However, if they use this new
eighth site as the source of initial excitation (which may be the
case, as it is located close to the antenna), they find that no
coherent exciton dynamics occur; only exponential decay. In
addition, they find discrepancies in the overall transport
efficiency, depending on which energies and couplings (i.e., which
Hamiltonian)  are chosen from the literature. In contrast, Olbrich
\textit{et al.} \cite{Olbrich11-2} used a Molecular Dynamics
approach to derive a time-dependent equation of motion for the
excitation transfer, and found that the eighth BChl did not form a
stable complex with the other seven, and that, while thermal
effects in total were stronger than assumed in other studies, the
coherent effects were enhanced when considering the trimer system
in totality.  This is attributed to a corresponding increase of
the site coupling within a monomer due to the influence of the
other monomers in the trimer.  However, they found that the
quantum coherent effects were still quite small in their model,
and Olbrich \textit{et al.} \cite{Olbrich11-2} argue that the fast
thermal fluctuations of the site energies are the main
contributing factor to the efficient transport of energy.

Inevitably a broader picture is needed before we can fully
understand the role of quantum coherence in natural conditions. Is
the coherence observed in experiments contributing significantly
to the efficient transfer of energy in natural conditions? {\em
In-vitro} experiments and theoretical models suggest a positive
answer, but ultimately experiments into other types and classes of
LHC systems (e.g., LH-1 and LH-2) are required, and a broader
understanding of the transport process {\em in vivo} is needed.

\section{Avian Magnetoreception: a tale of two spins}

Magnetoreception is a unique animal ability to detect either the
polarity or inclination of the earths magnetic field as a
navigation tool. As with electro-reception (the ability to detect
electrical fields, e.g., sharks hunting their prey), it represents
a sense slightly alien to human experience. Or so it was thought,
until recently. Experiments on manipulating a specific gene in
flies and monarch butterflies have indicated this gene plays a
significant role in magnetoreception, and since humans share this
gene in a dormant form it suggests that even we may harbor this
unique ability \cite{Foley10, Gegear}. However, some
evidence~\cite{Gegear} suggests it is not driven by the
radical-pair mechanism we discuss here.

The epitomic example of magnetoreception however is in migrating
birds. They apparently use this powerful ability, alongside other
biological geo-mapping tools, to navigate large distances. A
complete and satisfying description of the mechanism for the
magnetoreception ability of many species remains a mystery. Some
evidence suggests that a magnetite-based (deposits of magnetic
iron minerals) mechanism describes well this ability in some
animals \cite{WiltLight} (though recent reports suggest that, at least in pigeons,
these deposits are in fact macrophages and play no role in the magnetic sense \cite{Treiber12}). In other avian species,
particularly robins, several intriguing aspects of their behavior
rule out magnetite-based mechanisms. Here we focus on the
radical-pair model for magnetoreception that represents another
potential functional quantum biological system.  Precise details
will be given later, but summary radical pair is a pair of
molecules bound together, with each has an unpaired electron.
These electrons form naturally into singlet or triplet states, the
state of which, in turn, affects the disassociation, or reaction
products, of these radical pairs. It is proposed that the relative
weights of these reaction products which are biologically
detectable.  Thus, if the relative probabilities of having singlet
or triplet pairs is effected by the Earth's magnetic field, this
in turn affects the distributions of the two different reaction
products.  As a result, a biologically detectable signal can be
generated to function as a compass.

\begin{figure}[]
\includegraphics[width=0.9\columnwidth]{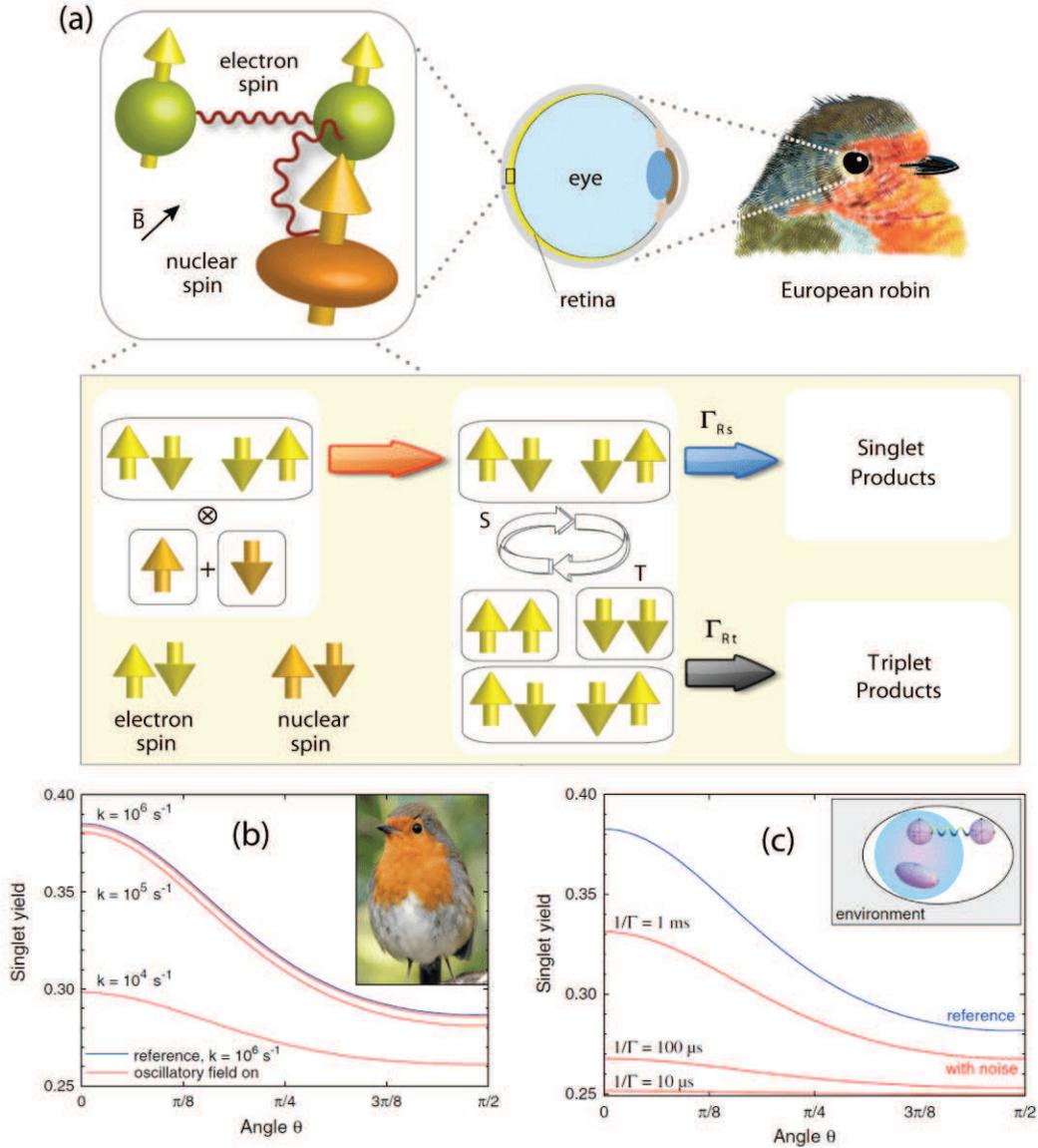}
\caption{{} (Color online) (a) The radical-pair reaction mechanism
as proposed for the avian compass. There are three main steps in
the usual process of magnetoreception via this mechanism: (1)
Light-induced electron transfer from one radical pair-forming
molecule (e.g., in the cryptochrome in the retina of a bird) to an
acceptor molecule creates a radical pair. (2) Singlet (S) and
triplet (T) electron-spin states inter-convert due to the the
external (Zeeman) and internal (hyperfine) magnetic couplings. (3)
Singlet and triplet radical pairs recombine into singlet and
triplet products respectively, which are biologically detectable.
(b) Singlet yield as a function of external field angle $\theta$
in the presence of an oscillatory field, but no noise (from
\cite{vedral11}). The blue curve shows just a static geomagnetic
field ($B_{0}= 47$ $\mu$T), and the red curves show the singlet
yield in the case where a $150$ nT field oscillating at $1.316$
MHz is superimposed perpendicular to the direction of the static
field.  An appreciable effect on the singlet yield occurs once $k$
(the decay rate of the radical, equivalent to $\Gamma_R$ as
defined in equation \ref{gammar}) is of order $10^{-4}$ $s^{-1}$.
(c) Singlet yield as a function of magnetic field angle $\theta$
for differing noise magnitudes (from \cite{vedral11}) (for
$k=10^{4}$). The blue curve shows the optimal case with no-noise.
The red curves indicate that a noise rate of $ \Gamma>0.1k$ has a
detrimental effect on the singlet yield.  Both of these results
indicate that the electron spin state must have a remarkably long
coherence (or correlation) time.\label{RP}}
\end{figure}

\subsection{Behavioral studies}

The hypothesis that migrating birds can make use of the earth's
magnetic field for orientation was first proposed by von
Middendorff \cite%
{Middendorff}. Later, European robins were found to utilize a
magnetic compass when migrating \cite{Wiltschko66}. Since then,
many experimental tests on the behavior of migrating birds have
been performed. In particular, it was found that for some species
their
magnetic sense acts as an inclination compass, but is insensitive to parity~\cite%
{wilt72}. In other words, they can detect the relative angle of
the local earth's magnetic field lines, but not whether they are
going north or south. There is some benefit to this in that the
compass would not be effected by geomagnetic
reversal~\cite{Wiltschko66}. Typically, however, most models of
magnetite-based magnetoreception act as a parity compass
\cite{WiltLight}.

 With advances in technology and technique, many more delicate behaviorial tests have been performed. For
example, it was found that the compass was: (a) dependent on the
frequency and intensity of ambient
light~\cite{Wiltschko00,WiltLight}; (b) dependent on the intensity
of the external magnetic field \cite{WiltLight}; (c) disrupted
when repeated magnetic pulses were applied to the migratory birds
\cite{Wiltschko02a}, and (d) disrupted when an external
oscillating field was applied at certain angles to the geomagnetic
field \cite{Ritz04}.  In addition, lateralization in terms of the
the birds' visual system associated with magnetoreception has also
been reported \cite{Wiltschko02}, suggesting the mechanism
operates primarily in the right eye.

All of this data lends support to a photo-activated radical-pair
mechanism. In particular, the experiments showing (d), the
disruption of the navigation mechanism by an external oscillating
field in the megahertz range (as seen in migratory robins and
directionally-trained chickens as well as zebra finches
\cite{WiltLight, Ritz04,Ritz09, Wiltschko07}), match exactly what
is predicted by simple models of radical-pairs, which we will
describe shortly.

Very recent evidence (see Ref. \cite{WiltLight} for a detailed
review) suggests that both an inclination compass and a polarity
compass are present in certain avian species, suggesting that a
magnetite and radical-pair compass can coexist.  We also point
out, as mentioned earlier, that in some species of flies,
butterflies, and even perhaps humans, there may be a third unknown
mechanism at work \cite{Foley10, Gegear} that has some of the same
behavioral properties of the radical-pair model, but not others.
Here we will focus on describing the evidence for the radical-pair
mechanism, and in what way that mechanism represents ``functional
quantum biology''.

\subsection{Radical-pair model: evidence and motivation}

Based on  findings of behavioral studies, it is obvious that a
successful model should contain a light-dependent factor. Another
feature is that because the geomagnetic field is very weak ($50$
$\mu$T), the magnetic compass mechanism must be very sensitive. In
addition, this mechanism must transduce the response to the
Earth's field into a biologically-detectable signal. Finally, as
we mentioned earlier, in some species the compass is only an
inclination compass.

To help satisfy some of these criteria, the radical-pair mechanism
(RPM) for magnetoreception was proposed by Schulten \textit{et
al.}~\cite{Schulten} in 1978. Within the radical-pair model there
exists the possibility of a photo-sensitive element \cite{Ritz00,
Ritz10}, i.e., absorption of light triggers an electron transfer
from a donor to an acceptor molecule to form the initial radical
pair (biologically this is proposed to occur within the
cryptochrome proteins within the eye).  This process can be made
to satisfy most of the frequency and intensity dependencies
observed in experiments on Robins \cite{WiltLight}.

How then is this donor-accepter system sensitive to the Earth's
field?  Most known radicals contain atoms (hydrogen and nitrogen)
with nuclear spins which act like an effective internal magnetic
field affecting the electron spin via the nuclear hyperfine
interaction. To sense the weak magnetic field of the Earth, the
combination of anisotropic internal nuclear hyperfine fields and external
geomagnetic fields causes a mixing of the electron singlet and
triplet states, the strength of which is dependent on the angle of
the external geomagnetic field.  Ritz et al ~\cite{Ritz10} assumed
a simplistic internal magnetic environment with only one
anisotropic nucleus (i.e., one radical is devoid of internal
magnetic fields, and the other radical has very strong internal
magnetic fields) and showed that this maximizes the
singlet-triplet mixing compared with other designs \cite{Ritz10,
Christopher09}.  Whether this occurs in nature is unknown, but
their extreme example helps to present a simple model of how this
mechanism works (see later).

For the last stage, as mentioned, this radical pair can decay into
different reaction products at different rates \cite{Ritz00,
Rodgers09} depending on the spin state (singlet or triplet).  If
this radical-pair is located in the cryptochrome proteins within
the eye, this could lead to a vision-based compass where the bird
actually sees a directional signal~\cite{Ritz00}.  Finally, this
model requires that the many molecules forming the
radical-pair-based magnetic compass must be ordered in some way so
that the directional effects will not be averaged out
\cite{Hill10, Ilia10}.  This may require that the radical-pairs
exist in a regular ordered biological structure, or lattice, which
has not yet been identified.

\subsubsection{The Schulten model}

A simple model to calculate the influence of the Earth's field on
the reaction products of the radical pair was outlined originally
by Schulten \cite{Schulten,Ritz00}.  Here we present specific
details of this model, to help clarify the physical picture
described in the previous section. As discussed above, the radical
pair is formed when an electron is transferred from a donor to an
acceptor molecule, typically via an optically-excited process.
This forms a singlet state for the combined donor-acceptor system.
The, ideally the subsequent dynamics of this state is primarily
affected by an anisotropic hyperfine interaction between the
electron spin and the nuclear spin within the molecules.  This
must occur faster than any decay (recombination) or decoherence of
the radical-pair. The interaction is described by a hyperfine
coupling Hamiltonian (assuming only one of the two spins is
affected by its internal nuclear spin, for the most sensitive
possible compass~\cite{Ritz10})
\begin{equation}
H_{hyp}= \hat{S}_{1}\cdot \bold{A} \cdot \hat{I},
\end{equation}
where $\hat{S}_{1}=(\sigma_{x}/2,\sigma_{y}/2,\sigma_{z}/2)$ is
the operator of electron spin $1$, $\hat{I}$ is the nuclear spin
operator, and $\bold{A}=(A_{x},A_{y},A_{z})$ in its diagonal basis
is the anisotropic hyperfine coupling tensor for a cigar-shaped
molecule with $A_{x}=A_{y}=A_{z}/2$ and $A_{z}=2.418$ MHz, for
high sensitivity \cite{vedral11} (though recent results have
suggested the weak-hyperfine coupling regime can produce a
stronger sensitivity, given the right combination of parameters
\cite{Cai12}). The Zeeman splitting of the electron spins is
 \begin{equation}
H_{B}= g\mu_{B}\bold{B}\cdot(\hat{S}_{1}+\hat{S}_{2}),
\end{equation}
where $\bold{B}$ is the magnetic field vector describing the
external geomagnetic field (or any other applied fields), $g=2$,
and $\mu_{B}$ is the Bohr magneton of the electron. The magnitude
of the geomagnetic field is typically of the order $B_{0}\approx
0.5$ G ($\approx 50$ $\mu$T). For the axial symmetry of the
hyperfine tensor, the magnetic field vector is explicitly written
as $\bold{B}=B_{0}(\sin\theta,0,\cos\theta)$, for
$\theta\in[0,\pi/2]$.

In addition to the coherent evolution driven by $H_{hyp}$ and
$H_{B}$, the radical-pair dynamics feels several incoherent
effects.  Both singlet ($S$) and triplet ($T_{+},T_{0},T_{-}$)
states undergo spin-selective relaxation or recombination into the
singlet product state $\left|\textbf{s}\right\rangle$ and the
triplet product state $\left|\textbf{t}\right\rangle$.  This
occurs when, e.g., there is electron back-transfer from the
acceptor radical to the donor radical, which is described by the
self-energy
\begin{equation}
\Sigma_{\text{p}}[\rho]=-\Gamma_{R}\sum_{\alpha,\beta}[s_{\alpha\beta}s_{\alpha\beta}^{\dag}\rho-2s_{\alpha\beta}^{\dag}\rho
s_{\alpha\beta}+\rho
s_{\alpha\beta}s_{\alpha\beta}^{\dag}],\label{gammar}
\end{equation}
where $s_{1\beta}=\left|S,\beta\right\rangle\,\left\langle
\textbf{s}\right|$,
$s_{2\beta}=\left|T_{+},\beta\right\rangle\,\left\langle
\textbf{t}\right|$,
$s_{3\beta}=\left|T_{0},\beta\right\rangle\,\left\langle
\textbf{t}\right|$, and
$s_{4\beta}=\left|T_{-},\beta\right\rangle\,\left\langle
\textbf{t}\right|$, for the up and down nuclear spin states
$\beta=\uparrow,\downarrow$. Here we assume that the singlet and
triplet products have the same recombination rates
$\Gamma_{Rs}=\Gamma_{Rt}=\Gamma_{R}$.  The final incoherent term
is the environmental noise. The simplest noise model is where both
phase and amplitude of both electron spins are equally perturbed
with self-energy
\begin{equation}
\Sigma_{\text{n}}[\rho]=-\Gamma_{\text{n}}\sum_{k,j}[L_{kj}L_{kj}^{\dag}\rho-2L_{kj}^{\dag}\rho
L_{kj}+\rho L_{kj}L_{kj}^{\dag}],
\end{equation}
where $L_{1j}=\sigma_{x},L_{2j}=\sigma_{y},L_{3j}=\sigma_{z}$ for
the electron spins $j=1,2$.

The dynamics of the electron and nuclear spins are then determined
by the Hamiltonian $H=H_{hyp}+H_{B}$ together with the incoherent
contributions $\Sigma=\Sigma_{\text{s}}+\Sigma_{\text{n}}$ for the
generalized master equation: $\dot{\rho}(t)=\mathcal{L}[\rho(t)]$,
where the Liouvillian $\mathcal{L}$ is
$\mathcal{L}=-i[H,\cdot]+\Sigma$. 

The biologically-detectable signal is the integrated triplet or
singlet yield. For example, the triplet yield is, \beq \Phi(t) =
\int_0^{\infty} \Gamma_{Rt} T(t) dt,\eeq where, $T(t) =
\mathrm{Tr}[Q^T\rho(t)]$, and $Q^T$ is the projector onto the
triplet states.  The interplay between the external geomagnetic
field and the anisotropic nuclear hyperfine interaction results in
a dependence of the triplet yield on the relative angle between
the geomagnetic field and the radical-pair molecule.

This simple model is sufficient to describe the destructive effect
of additional radio frequency fields, as observed in experiments,
and enables one to estimate the most ideal properties of the
radical-pair system to maximize the sensitivity of the magnetic
compass.

\subsubsection{Radical-pair molecules: experimental tests}

As discussed earlier, a variety of behaviorial experiments support
the radical-pair mechanism; e.g., the disorientation of the
magnetoreception ability in Robins when external magnetic fields
of around a megahertz frequency are applied~\cite{Wiltschko07,
Keary09, Ritz09}. In addition, comparison of the
frequency-dependence of this behavior to theoretical models
indicates that the radical-pair system does have the optimal
design mentioned above; i.e., one of the radicals may contain
almost no nuclear spin polarization.

To further bolster evidence supporting the radical-pair model,
there are also experimental tests at the molecular level on a
variety of candidate molecules that could be at the heart of the
mechanism.  There have been a large amount of studies on
radical-pairs in solution, and on the effect of external fields on
the reaction products (see a review of early experiments by
Steiner and Ulrich \cite{Steiner89}) though only under magnetic
field strengths much stronger than Earth's. In recent work,
Woodward et al \cite{woodward01} developed a technique for
studying the properties of such candidate reaction-pair systems.
As an example they studied the photochemical reaction of pyrene
(Py) and N,N-dimethylaniline (DMA) in solution under a very weak
(500 $\mu$T) radio-frequency magnetic field. When pyrene is
continuously irradiated by UV light, in the presence of DMA, the
transient radical ion pair Py$^{\cdot -}$ DMA$^{\cdot -}$ is
produced in a spin-correlated singlet state and is found to
interconvert with the corresponding triplet state, as expected. An
exciplex (an electronically-excited complex of Py$^{\cdot -}$ and
DMA$^{\cdot -} $) is then produced due to the electron-hole
recombination of the singlet radical pair. Their technique uses
the fluorescence of the exciplex to monitor the concentration of
singlet pairs, and estimate the reaction products. They found that
even this very weak magnetic field induced changes of up $25\%$ in
the reaction products.

 Rodgers \textit{et al.} \cite{Christopher07} also studied the spin-selective recombination of Py$^{\cdot -}$ and DMA$^{\cdot -}$ radicals under the affect of $0$--$23$ mT
magnetic fields. It was found that weak magnetic field effects
become most pronounced when the ratio of hyperfine coupling
strengths for the two radicals is large. Another recent example is
the triad composed of linked carotenoid (C), porphyrin (P) and
fullerene (F) groups considered by Maeda \textit{et
al.}~\cite{Maeda08}. By using spectroscopic measurements they
found this particular radical-pair's lifetime was changed by the
application of magnetic fields less than $50$ $\mu$T (as needed to
detect the Earth's weak geomagnetic field). In addition, they
found this effect responded
anisotropically 
to the angle $\theta$ between the molecule and the external field,
which is of course an essential feature of a radical-pair compass.
However the anisotropy was only manifest at stronger fields ($3.1$
mT), and no anisotropy could be observed for $50$ $\mu $T, even at
low temperatures ($113$ K).  It is hoped that future work will
show whether a radical pair formed by ``reduced flavin cofactor
and an oxidized tryptophan residue in a cryptochrome
flavoprotein'' (Maeda et al~\cite{Maeda08}) will have the long
spin coherence lifetime and low recombination time to achieve the
$50$ $\mu$T sensitivity at room temperature needed for this
mechanism to function.

\subsection{Functionality and significance of quantum coherence}

Attention has recently focused on the functionality of quantum
coherence in the radical-pair model because, as we
noted in the introduction, spin singlet and triplet states are
inherently quantum.  This has led to a great degree of interest in
this phenomena from researchers in condensed matter and quantum
information theory.  
For example, an analysis of the entanglement and decoherence
effects of this mechanism \cite{vedral11} suggests that, in their
words, for this model to function as desired, ``{\em superposition
and entanglement are sustained in this living system for at least
tens of microseconds, exceeding the durations achieved in the best
comparable man-made molecular systems}''

This implies that the radical-pair model is, if it really exists
and functions as described, a piece of functional quantum
biological hardware. From the perspective of condensed matter
physicists, it is very akin to man-made artificial molecule
systems, e.g., the singlet and triplet states of spin-blockaded
electrons in double quantum dots \cite{Kouwenhoven}.  However,
further work remains to be done on identifying the exact radical
pair which sustains this mechanism, to show that it has the
properties necessary to act as a compass, and to verify this with
further {\em in vitro} experiments.

\begin{figure}[]
\includegraphics[width=0.7\columnwidth]{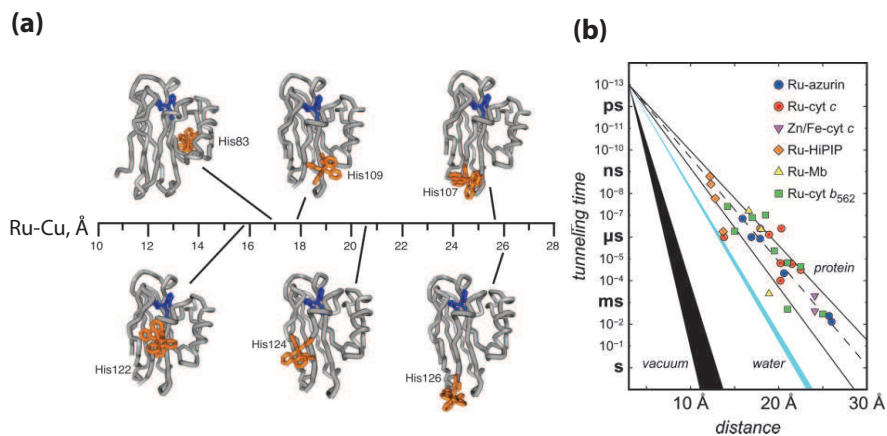}
\caption{\label{protein_et} (Color online) (a) Tunnelling
distances
 (Ru-Cu) and backbone structure models of several Ru-modified
 azurins. The azurin protein has a Cu active site (blue) and several
 His residues that can be modified to attach a
 Ru(bpy)$_2$(im)(HisX)$^{2+}$ label (orange). This allows
 measurements of Cu(I)$\rightarrow$Ru(III) electron transfer at
 various distances through the protein $\beta-$sheet. (b) Tunnelling
 timetable for intraprotein electron tunnelling in
 Ru-modified proteins. In general, the distance dependence shows an
 exponential decay, illustrating a single-step tunnelling
 mechanism. The two solid lines depict the
 predictions for coupling along $\beta$-strands (decay constant
 $\beta=1.0 \AA^{-1}$) and $\alpha$-helices ($\beta=1.3 \AA^{-1}$)
 predicted by the tunnelling-pathway approach. The dashed line shows a
 $\beta=1.1 \AA^{-1}$ distance decay in  proteins. Distance dependence
 for electron tunnelling through water as well as vacuum are also
 shown. Both images are from Ref. \cite{Gray:2005p79841})}
\end{figure}


\begin{figure}[]
\includegraphics[width=0.7\columnwidth]{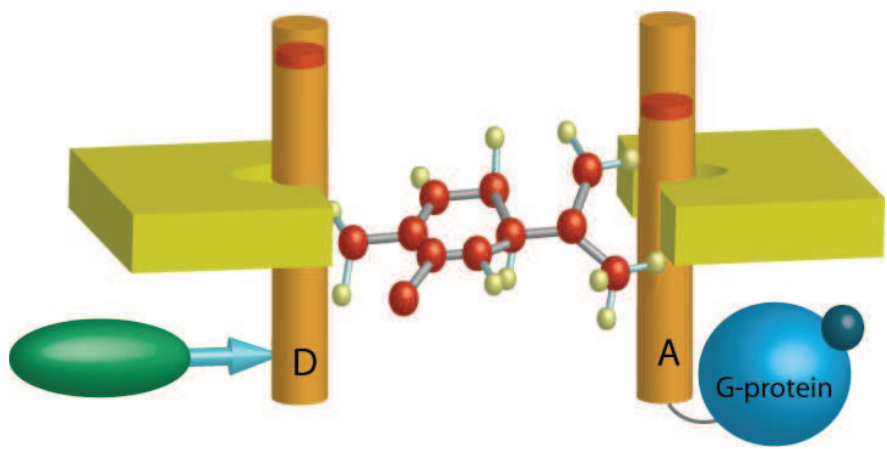}
\caption{{} (Color online) The schematic shown above outlines the
model proposed by Brookes et al \cite{Brookes07} to describe
charge-transport-mediated odorant discrimination. In general, the
olfactory receptor ($G$, a transmembrane protein) responds to an
odorant molecule by releasing a subunit of a neighboring $G$
protein, starting a chain reaction of Ca ion influx into the cell
(see the discussion of ion channels for how this process may also
have quantum features). In the model proposed by Brookes et al
\cite{Brookes07}, an electron source $X$ arrives and exchanges
charge with the receptor protein. The charge travels to a donor
$D$ in one part (helix) of the transmembrane receptor protein. The
charge can then hop, or tunnel, to the acceptor A, in another
helix of the protein, inelastically assisted by an odorant phonon
emission. The charge, after arriving at $D$, triggers the release
of the neighboring $G$ protein subunit. Many precise features of
this model are unknown. For example, the origin of the electron
source $X$ is unknown, but is conjectured to be an oxidizing agent
in the cell fluid, which diffuses throughout the cell and randomly
arrives at the transmembrane to initiate the charge-transport
process.  The goal of this model is to have the release of the
neighboring $G$ protein subunit depend on vibrational properties
of the odorant molecule, and thus discriminate different molecules
beyond features like size and shape.\label{smell}}
\end{figure}

\section{Other possible quantum biological systems}

\subsection{Tunneling in biological systems}

In many biological processes, tunneling of light particles
(hydrogen or electrons) provides an alternative route to classical
over-the-barrier reactions, resulting in new channels for chemical
reactions that are forbidden in classical mechanics. Tunneling in
biological systems was first reported in long-range
electron-transfer processes in proteins
\cite{Gray:2003p83392,Stuchebrukhov:2003p83910,Gray:2005p79841},
and then in the transfer of a proton, hydride, or hydrogen atom in
various enzymatic catalytic reactions
\cite{Allemann:2009uq,Nagel:2006p82823}. Furthermore, the
simultaneous transfer of an electron and a proton from different
sites (so-called proton-coupled electron transfer) also play an
important role in many biological functions
\cite{HammesSchiffer:2010p91583}. A significant increase in the
enzyme catalytic efficiency or long-range electron transfer
through proteins due to tunneling have been observed in many
biological systems. This class of quantum phenomena in biology
clearly fits into our description of functional quantum biology.

Electron transfer between redox centers separated by distances of
the order of 15-30 $\AA$ plays important roles in respiration and
photosynthesis \cite{Gray:2003p83392,Stuchebrukhov:2010ex}. In the
past three decades, Gray and coworkers
\cite{Gray:2003p83392,Gray:2010p83224} have yielded a remarkably
detailed description of the distance- and driving-force
dependencies of long-range electron tunneling rates in
ruthenium-modified proteins. Particularly, Ru-azurin has proven to
be an excellent model system for electron tunneling through folded
proteins (Fig. 7a) \cite{Langen:1995p83554}. Experimentally
observed electron transfer rates that exhibit exponential decay as
distance increases and weak temperature dependence indicate a
single-step tunneling mechanism (Fig. 7b). It is remarkable that
such long-distance electron transfer in biology occurs via quantum
mechanical tunneling. Tunneling over such long distances would be
impossible in vacuum, however, through bonded as well as
non-bonded interactions in the protein, electron tunneling
barriers can be significantly reduced, leading to significantly
higher electron tunneling rates in proteins.

A major
open question remains in this field is whether or not
the electron conducting proteins have evolved specific pathways for
electron tunneling
\cite{Stuchebrukhov:2010ex,Gray:2003p83392,Moser:2010p83703}.
Theoretical analysis has demonstrated that an empirical model treating
the protein as a structureless random medium explains the experimental
data \cite{Moser:2010p83703,Page:1999p83650}. However, structure-based
analysis reveals that specific channels through covalent bond, hydrogen
bond or even van der Waals contacts form electron tunneling
{\it pathways} that facilitate electron tunneling over a long distance
\cite{Onuchic:1992p83588,Regan:1999p69489,Gray:2003p83392,Stuchebrukhov:2003p83910,Skourtis:2010p83621}. Are
these redox proteins specifically wired to perform efficient electron
tunneling? How much do the functions of these biological systems
depend on the quantumness of the process? Again, these are not
questions that can be answered easily, and interpretation of current
experimental results often involves ambiguities.

Recently, Skourtis and coworkers have suggested a proposal that is
likely to provide an incisive test for the pathways model. Quantum
theory predicts that electron transfer pathways could interfere
with each other \cite{Regan:1999p69489}, and it has been suggested
that electron transfer through the azurin protein depends
critically by quantum interferences between multiple distinct
pathways \cite{Regan:1999p69489}. Motivated by these earlier
theoretical studies, Skourtis and coworkers suggested a molecular
which-way interferometer experiment in which localized and
distinguishable normal modes coupled to bridge atoms can be
vibrationally excited to control inelastic scattering of electrons
and interferences between transfer pathways
\cite{Xiao:2009p83622,Skourtis:2010p83621}. Experiments have
demonstrated that electron transfer in a a small
donor-bridge-acceptor organic molecule can be controlled by
excitation of vibrational modes localized on the bridge moiety
\cite{Lin:2009p82424}. Such experimental probes of pathway
coherence in electron-transfer proteins in the single-protein
level would provide decisive proof for the quantumness and
functionality of the electron tunneling process.

Tunneling of a proton, hydride, or hydrogen atom also play an
important role in a wide range of biological enzymatic catalytic
reactions \cite{Allemann:2009uq,Nagel:2006p82823}. Measurements of the
intrinsic kinetic isotope effects (KIE) in enzymatic reactions have
clearly demonstrated nuclear quantum effects in enzymes
\cite{Allemann:2009uq,Nagel:2006p82823}. For example, in soybean lipoxygenase
\cite{Rickert:1999p85314} and methane mono-oxygenase
\cite{Nesheim:1996p85318}, large KIEs with $k_H/k_D$ close to 100 have been
observed, and the sizes of the KIEs clearly indicate quantum-tunneling
effects. In enzyme catalysis, a large portion of the quantum
improvement over the classical catalytic rates
can be attributed to the energy shift due to the zero-point energy
that gives a quantum correction to the barrier height and the
H-tunneling rates \cite{Allemann:2009uq}.
Note that semi-classical models including environmentally coupled H-tunneling
have been shown to adequately describe H-tunneling in
enzymes \cite{Marcus:2007p6254,HammesSchiffer:2008p57771}.  Such nuclear
quantum effects in enzymes might represent a class of quantum phenomena in
biological systems that depends only on trivial quantum effects, not
quantum coherence.

\subsection{Smell}

Our sense of smell allows us to discriminate between small
molecules in very low concentrations via scent molecules
interacting with receptors in the nose. At this time, the
biomolecular processes of olfaction are not fully understood, and
some evidence suggests that a mechanism based solely on the size
and shape of odorant molecules is inadequate.  For example, it has
been noted that molecules with very similar shapes and sizes have
a remarkably different scent~\cite{turin96}. Thus traditional
models of a ``docking''-type mechanism, where the  size and shape
of an odorant molecule actuates the receptor in some way, are
thought to be insufficient. Turin \cite{turin96} proposed a
mechanism which, in addition to ``docking'', gives a further level
of selectivity (or sensitivity) by a process of inelastic electron
tunnelling. In this case the odorant molecule both docks with a
receptor and then mediates phonon-assisted inelastic tunnelling of
an electron from a donor to an acceptor (i.e., donor and acceptor
electronic states differ in energy by $\hbar \omega $, and thus
transport only occurs when energy is conserved by emission of an
odorant phonon, the vibrational degree of freedom of the molecule
one is ``smelling'', of the right energy).

A recent model proposed by Brookes~\cite{Brookes07} expanded on
this idea, and presented evidence that such a mechanism fits the
observed features of smell, and is at least ``physically''
credible (see Fig.~[\ref{smell}] for an overview and details of
the mechanism). Whether such a mechanism ultimately exists in
nature has yet to be determined. While not specifically requiring
``coherence'' to function, this mechanism requires inelastic
phonon-assisted tunnelling of electrons, and is certainly more
``microscopic'', and sensitive, than previously thought.

\subsection{Quantum coherence in ion channels}

Ion channels, which regulate the flow of ions across the membrane
of a cell, are a vital component of many biological processes
\cite{Hille01}, e.g., neuronal communication (potassium channels),
muscle contraction, etc. In many cases the flow of ions through
the channel is controlled by a gate that can be activated by a
voltage, a chemical signal, incident light or mechanical stress.
In addition the channels often feature a filter which controls
which types of ions may pass.  This selectivity filter is only a
few angstroms wide, which forces the ions to pass through it
one-by-one~\cite{Roux04}.

A recent conjecture by Vaziri and Plenio \cite{Vaziri10}
is that the selectivity filter of ion channels,
and the transport of ions through the channel, may exhibit quantum
coherence.  They are motivated by the large selectivity found in
the bacterial KcsA potassium channel, which can select potassium
over sodium with a ratio of $10^4$. They argue that the energy
scale of the thermal fluctuations of the atoms which form the
channel suggest two possible quantum coherent phenomena. Firstly,
there could be a coherent diffraction of a 1-D potassium ``matter
wave'' off an effective grating formed by modulations of the
potential energy landscape inside the channel. Secondly, since the
channel forms a 1-D array of trapping sites, the potassium ions
could take advantage of quantum tunnelling to process through the
channel in an efficient manner. They calculate a tunnelling rate
which is of the same order as the decoherence time, and thus argue
that an ``environment assisted'' quantum tunnelling phenomena,
akin to that seen the energy transport in photosynthesis, is
responsible for the high selectivity and throughput of these
channels. However, unlike in the transport of energy in FMO, this
process involves the movement of charges through a complex
environment, with a correspondingly large re-organization energy,
which implies the coupling to the environment may be too strong
for coherence to thrive.

This conjecture is intriguing, but at this time lacks experimental
support (apart from the already observed remarkable performance of
the channels).  Vaziri and Plenio \cite{Vaziri10} propose several
ways to verify the presence of this ``functional quantum effect'',
focusing on collective transport resonances which, they argue, are
the functionality provided by the quantum coherent transport in
this case. Furthermore, a nitrogen-vacancy (NV) probe in proximity
to the ion channel \cite{Hall10} might be a promising tool to
observe any possible quantum coherent effects.

\subsection{Photoreceptors}

Biological photoreceptors, i.e. light-sensing proteins, are ubiquitous
in the functions of life forms on Earth. These proteins contain
chromophores that upon excitation, undergo ultrafast chemical transformation on
their excited-state potential energy surface and eventually lead to
light-induced signal transduction
\cite{vanderHorst:2004cc,Sundstrom:2008p27180}. Because of the quantum nature of
the electronically excited state and nuclear quantum coherence that
often accompany excited-state molecular dynamics in the femtosecond
time scale, quantum mechanics are required to describe these
photo-activated dynamics in biological photoreceptors~\cite{Onuchic:1988vn}.

\begin{figure}[]
\includegraphics[width=0.7\columnwidth]{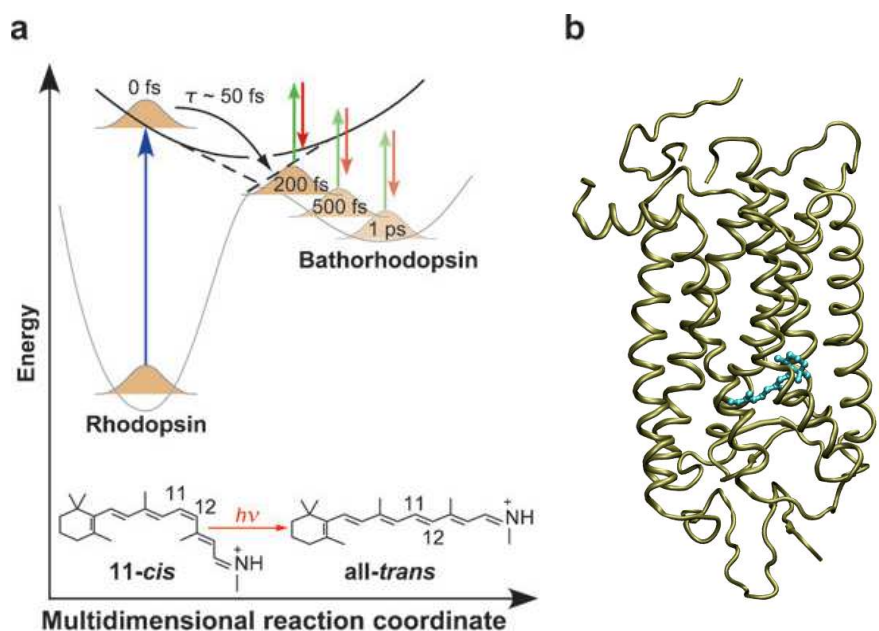}
\caption{\label{rhodopsin} (Color online) (a) Schematic potential energy surfaces of
  electronic states involved in the photoisomerization of retinal
  (image from Ref. \cite{Kukura:2007p14138}). (b) Structure model of bovine
  rhodopsin. The retinal motif is rendered in cyan. The quantum yield
  of $\sim65\%$ is almost an order of magnitude higher than that
  observed in the absence of the protein environment.}
\end{figure}

For example, the primary event in vision involves the
photoisomerization of a retinal molecule in the trans-membrane
protein called rhodopsin (Fig. \ref{rhodopsin}), which is the most
well studied and characterized photoreceptor. Following light
absorption, this photoisomerization reaction proceeds with
remarkable rate ($<$200 fs, one of the fastest chemical reactions
known) and high specificity (quantum yield of $\sim0.65$)
~\cite{Schoenlein:1991to}. This isomerization triggers a protein
conformational change that is subsequently amplified through
protein-protein interactions with signal-transduction proteins,
leading to the visual signal. The efficient
photoisomerization process makes the retinal molecule an effective
and reliable photo-switch. The quantum mechanical positioning of
electronic states and their symmetries enables ultrafast coherent
wave-packet dynamics that are responsible for the rapid
photoisomerization reaction
\cite{vanderHorst:2004cc,Sundstrom:2008p27180,BenNun:1998vx}. In
addition, the unique reactivity of rhodopsin has often been
attributed to a conical intersection between its electronic
excited state and ground state \cite{Abe:2005ea,Garavelli:1997ue},
which has recently been demonstrated experimentally
\cite{Polli:2010bk}. Nuclear quantum coherence clearly plays an
important role in the remarkably rapid and efficient
photo-activated one-way reaction dynamics in rhodopsin
\cite{BenNun:1998vx}. Intriguingly, coherent optical control of
retinal isomerization yields in bacteriorhodopsin, a protein
closely related to vertebrate rhodopsins, has been demonstrated
experimentally \cite{Prokhorenko:2006cg,Prokhorenko:2011co}. The
frontiers of the investigations of biological photoreceptors are
thus to further elucidate the quantum nature that controls the
high yield and specificity of natural photoreceptors in order to
design and direct chemical dynamics using coherent light
\cite{Nagy:2006bs}.

\section{Open issues}

\subsection{Open problems in photosynthetic light harvesting}

\subsubsection{Protein Environment and Spectral Density}

The protein environments surrounding pigments in photosynthetic
complexes play crucial roles in light harvesting~\cite{Gilmore:2008fn,Ishizaki:2010p82121}. On the one hand,
the protein scaffolding provides a rigid structure that fixes
photosynthetic pigments in space to form an antenna network,
and pigment-protein interactions also shift the transition
energies of individual pigments to yield the energy landscape that
directs energy flow towards the reaction
center~\cite{Adolphs06,Renger:2009p77537}.
These aspects of proteins' roles are static and well recognized, and
as a result structure-based studies
have formed a large component in this research field~\cite{Warshel:2001uy}.

On the other hand, dynamical fluctuations of pigment-protein couplings
are directly responsible for the energy transfer
dynamics. However, these dynamical effects are often overlooked
and treated without details, as the simple models of spectral densities
described in Sec. 2.2. Does
the popular diagonal exciton-phonon coupling model completely
capture the physical reality in photosynthetic complexes? How
important (or unimportant) are off-diagonal exciton-phonon
couplings? Are the simplified spectral density functions used in
the literature adequate to describe the real system? Can
correlated baths significantly influence exciton dynamics? And if
yes, can the correlated-bath effects be directly observed in real
systems? What are the importance of non-Markovian effects? Can we
directly measure non-Markovianity and the memory kernel of system-bath
interactions? All these questions are still unresolved and require
a combination of experimental and theoretical studies in order to
provide adequate answers. Experimentally, probes more specific to
exciton-phonon interactions and bath dynamics must be developed.
In this regard, the rich information in the peak shapes and
spectral dynamics of 2D electronic spectra must not be overlooked
\cite{Ginsberg:2009p76649,cheng2009dynamics,Abramavicius:2009p75282,Chen:2011bm}.
Theoretically, simulations of exciton transfer dynamics in
photosynthetic systems must be expanded to investigate more
general models including different types of exciton-phonon
couplings and bath spectral densities~\cite{Ishizaki:2010p82121}.
Recently, molecular dynamics simulations combined with quantum
chemistry calculations have been applied to study fluctuations in
the FMO complex \cite{Olbrich11,Olbrich11-2} and reaction center \cite{Damjanovic:2002p884,Jing:2012kb}. This type of study
might lead to new insights on the dynamical role of the protein
environments and provide atomistic resolution to the spectral
densities; however, the interpretation of the
hybrid-quantum/classical results should be treated with caution
\cite{Zwier:2007p328,Damjanovic:2002p884}.

\subsubsection{Consequences of Coherence}

As we have discussed in Sec. 2, quantum coherence effects in
photosynthesis have drawn intense research interest recently. So
far, experimental and theoretical works have focused on the
question of how quantum coherence effects contribute to improve
the efficiency of energy transfer, and a measure of efficiency
such as the quantum yield defined in Eq.~(4) is often adopted as
the target to benchmark the effects of quantum coherence. It is
clear that site-basis quantum coherence, i.e. coherent
delocalization of excitons, plays crucial roles in determining the
energy landscape and spectral properties of photosynthetic
complexes \cite{Yen:2011p99898,Fleming:2011p99897,Scholes:2011kx,Struempfer:2012ep,Ishizaki:2012kf}.
However, the consequences of excitonic quantum coherence
responsible for the reversible wave-like exciton dynamics are less
clear. As we have detailed in Sec. 2.3, it has been speculated
that excitonic coherence can improve functions such as robustness
against energetic disorder or dynamical perturbations. In
addition, Ishizaki and Fleming \cite{Akihito09} argued that
excitonic coherence can be utilized to form energy ``ratchets''
for energy transfer that avoid trapping by local energy minima.
However, these results are often model-dependent and a unifying
view of excitonic coherence effects in photosynthesis seems to be
still lacking.

Furthermore, to establish excitonic coherence as a general and
essential phenomena in photosynthetic light harvesting,
measurements on larger photosynthetic complexes such as LH2, LH1,
and LHCII must be carried out to confirm the general existence of
long-lasting quantum coherence. It is also necessary to consider
more general initial conditions beyond the coherent
laser-excitation condition suitable for the interpretation of
laser experiments~\cite{Mancal:2010kc}. Does quantum coherence
play a role when excitations are generated by incoherent sunlight
or energy transfer from another antenna complex? Another open
question that follows naturally after the observations of
long-lasting quantum coherence in photosynthetic complexes is
whether or not quantum effects (quantum coherence and
entanglement) in photosynthetic systems can be used as a quantum
resource? Here, a crucial experimental development is the
capability to prepare and measure an excitation localized on a
single chlorophyll site, which would allow the quantum mechanical
manipulation of the wavefunction of excitons.

\subsubsection{Biological Relevance}

So far excitonic quantum coherence has been observed only in {\it in
vitro} experiments, and theoretical studies have only established that
quantum coherence could affect efficiency of energy transfer
within a single complex. However, a connection between excitonic
quantum coherence effects and biological function {\it in vivo}
has yet to be established~\cite{Gilmore:2008fn,Wolynes:2009hs,Hoyer:2010fl}.
In particular, the functionality of light harvesting
systems depends on assemblies with a length-scale beyond a single
complex; the whole antenna could contain tens of light-harvesting
complexes. Does the single-complex efficiency provide enough
information to determine the
energy transfer efficiency of the whole antenna? Moreover, what is
the system optimized for? It is conceivable that in the larger
length scales and longer time scales that determine the
functionality of photosynthetic light harvesting, biological
systems actually strive to achieve robustness and safety, instead
of maximal efficiency. It is therefore necessary to investigate and model
dynamics of light harvesting in larger assemblies including the
whole antenna. Specifically, experimental probes for the combined
temporal and spatial dynamics of exciton motions on the antenna
have to be developed \cite{Dawlaty:2011ft}. In addition, evidence of long-lasting quantum
coherence has yet to be obtained  {\it in vivo}. Theoretically, models of
quantum coherence effects should be considered for larger
structures, and it is necessary to develop measures of quantum
coherence that can be linked with {\it in vivo} observables that
can be determined and tested in experiments.

\subsubsection{Design Principles}

It is surprising that after decades of research we still do not
know how to build artificial light-harvesting systems that can
rival the efficiency achieved in natural photosynthesis
\cite{Scholes:2011kx,Fleming:2012et}. We know that: (1) chromophores with strong
absorbance must be used, (2) the antenna must contain high-density of
light absorbers yet avoiding the formation of exciton quenching
charge-transfer species, and (3) a energy gradient can help to funnel
excitations towards the reaction center \cite{Scholes:2011kx}.
However, the knowledge required to implement these principles in
artificial systems is lacking. Moreover, certainly there exist
other design principles in the working of natural photosynthesis.
If quantum coherence can help the efficiency of light harvesting,
then what are the molecular architectures that can utilize this
advantage to make a more efficient antenna? It is crucial that
theoretical studies focused on how chromophores should be arranged
are carried out and tested by experiments, maybe with the help of
synthetic organic chemists, to eventually formulate the design
principles that can help us build more efficient solar cells.

\subsection{Open problems in the radical-pair model}

\subsubsection{Radical Pair Candidate Molecules}

The discovery of cryptochromes as a possible candidate to be the
light-sensitive host of the magnetoreception radical pair
mechanism gives hope to the radical-pair model.  However, in-vitro
experiments have yet to show that cryptochromes have the exquisite
and anisotropic magnetic field sensitivity needed to function as a
magnetic compass.   Ideally cryptochromes, or other candidate
molecules, need to have some of the following
properties (see \cite{Rodgers09} and \cite{Ritz2011} for a more in-depth review of these issues):

\begin{itemize}
\item Spin-correlation and coherence time on the scale of
geomagnetic B-field effects. The rate of coherent spin oscillations induced
by the geomagnetic field must exceed that of any environmental
losses.  A $50$ $\mu$T field induces oscillations on the time
scale of $700$ ns.  This coherence, or correlation, must be
protected from environmental losses on this time scale for
significant change in the reaction rates to occur. Such
correlation times have been seen in
``photolyases'' radical pair systems.

\item Spin-correlation time and coherence on the same time-scales
as weak disruptive radio frequency fields.   In behavioral
experiments it was found that the magnetic sense was disrupted for
oscillating fields with very small magnitudes.  As with the
geomagnetic sensitivity this implies the spin correlation time
must survive for an exceptionally long time.  For example, in
Ref.~\cite{Ritz09}, Robins were found to be sensitive to fields as
weak as $50$ nT.  Ritz \cite{Ritz2011} discusses this point and
argues that the effect of a $10$ nT field induces a change in the
reaction yield of $0.1$\% at $10$ $\mu$s, and $1$\% at $100$
$\mu$s (a similar effect was shown in the calculation in
\cite{vedral11}). This implies both an exceptionally long spin
correlation time and extreme sensitivity of the magnetoreception
sense to changes in the reaction yield (see Fig.~(\ref{RP})).
\item Optimization of the hyperfine field.  Recent work by Cai {\textit et al} \cite{Cai12} suggests that a highly anisotropic but weak nuclear hyperfine coupling can give a stronger angular dependance on the yield than a strong one.  Whether this occurs in candidate radical pairs needs to be determined.
\end{itemize}

\subsubsection{Behavioral Tests}

There are a variety of further behavioral tests that have yet to
be done \cite{Rodgers09, Ritz2011}, and which might provide even
stronger proof for the radical pair model:

\begin{itemize}
\item Functionality window:   For static fields it was found that
changes in the intensity by more than 25\% disrupted the  magnetic
compass mechanism.  However, after a period of time the sense
returned.  Ritz \cite{Ritz2011} has raised the possibility that a
behavioral test for a similar ``re-adjustment over time'' to the
disruptive effect of oscillating fields will reveal if the
disruption is related to the functionality window seen with static
fields.

\item Magnetoreception via oscillating fields:  If the oscillating
field amplitude is of similar intensity to the Earth's geomagnetic
static field (in an environment where the geomagnetic field is
suppressed), some of the simple radical pair models predict that
the oscillating field with also induce a directionally sensitive
change in the reaction yields.  Rodgers et al \cite{Rodgers09}
argue that observation of  animals responding in a directional way
to this effect (essentially navigating by the oscillating field)
would be stronger proof of the radical pair mechanism than the
disruption effects observed so far (which in principle could be
attributed to some unknown causal effect on the motivation of the
animals).

\end{itemize}

Apart from these directly testable issues, there remains a lot to
be learned about the transduction of the radical pair yield to a
``compass'' that is biologically useful.  The chain of events
connecting changes in concentrations of molecules, into nerve
impulses, and then into a signal of directionality is completely
unknown. Indeed, as we have mentioned earlier, a large range of
information is probably combined to ultimately form the final
relevant directional cues (or instincts). This also has an impact
on an understanding of why very small changes in the radical pair
yields can apparently disrupt the magnetoreception sense. Finally,
there have been proposals for using quantum control techniques, in
the form of magnetic field pulses, to coherently control the spin
states \cite{Briegel10}.  If this induced a directional response
in test animals this would be extremely difficult to explain via
any alternative magnetoreception mechanism.  

Finally, a very recent experiment on pigeons has
identified a neural substrate responsible for processing magnetic sense information \cite{Wu12}. This substrate
was shown to respond to the intensity, direction, and polarity of the magnetic field, perhaps ruling out a radical pair mechanism in
this case.  For pigeons this is particularly interesting as other recent evidence \cite{Treiber12} has contradicted the previously held notion that magnetite deposits in their beaks were responsible for their magnetic sense.  Perhaps a third, unknown, mechanism plays a role in this particular species.  Similar experiments on neural response in species in which the radical pair model is thought to reside (e.g., european robins) are vital.

\section{Concluding remarks}

At the atomic level, the demesne of quantum mechanics, one can
always say that of course we exist in a quantum world (or at least
a non-classical one), and that life itself is thus, in some
manner, quantum. However, since the birth of quantum mechanics,
the extent to which the features of quantum mechanics, such as
quantum coherence or entanglement, can operate on biological time
and energy scales in a ``hot and wet'' environment, and provide a
biological advantage, has been seen mostly as a curiosity.

The recent development of quantum information theory has revealed
that quantum mechanics can, in some situations, do things better
than an equivalent classical theory (such as speed up computation
and provide secure communication).  Thus, as we have described
here, the goal of ``functional
quantum  biology'' is in some sense similar; to enhance the efficiency of transport in a molecular chain, or to
provide an exquisite sensor of geomagnetic fields.  A very extreme example is the idea proposed by Penrose \cite%
{Penrose}: that quantum computation may occur in microtubule
assemblies within the neurons of the brain. This idea has
attracted much attention, criticism and debate \cite{Max00,
Laura09} because it portrays the fascinating possibility of
quantum mechanical effects playing a central role in cognitive
function.  Evidence so far is lacking \cite{Max00, Laura09}, but
this idea has helped stir up a broader interest in the role of
quantum effects in biological processes, and the hunt for other
candidates of functional quantum biology.

Thus, at this time, the most promising candidates for function quantum biology remain photosynthetic units \cite%
{LeeScience07,EngelNature07,PNAS,ambientTemperature}, and
magnetoreception \cite{Ritz00,WiltLight}.  The evidence so far is
in favor of both these systems not only containing quantum
coherence in a hot and wet biological environment, but also that
it is used to gain a biological advantage.  There is a great
possibility that further examples of functional quantum biology
remain to be found.
 After all, as in a great number of technological breakthroughs, we usually find that
nature has been there before us.



\section{Acknowledgements} 

We wish to thank Jian Ma, A. Y. Smirnov and S. De Liberato for
helpful discussions. NL is supported by the RIKEN FPR Program.
YCC thanks the National Science Council, Taiwan (Grant No. NSC
100-2113-M-002-004-MY2), National Taiwan University (Grant No.
10R80912-5), and Center for Quantum Science and Engineering
(Subproject: 10R80914-1) for financial support.
 YN thanks the National Science Council, Taiwan (Grant No. NSC
 98-2112-M-006-002-MY3) for financial support. FN acknowledges partial
support from the NSA, LPS, ARO, NSF grant No. 0726909, JSPS-RFBR
contract No. 09-02-92114, MEXT Kakenhi on Quantum Cybernetics, and
the JSPS-FIRST Program.

\bibliographystyle{elsarticle-num}

\bibliography{bibliography,cheng}

\end{document}